%% file: main.tex
\documentclass[acmlarge]{acmart} 
\usepackage{xcolor}
\usepackage{color, colortbl}
\usepackage[letterpaper]{}
\usepackage{url}
\usepackage{booktabs}
\usepackage{multirow}

\usepackage{graphicx}
\usepackage{amsmath}
\usepackage{bm}
\usepackage{subcaption}
\usepackage{pbox}
\usepackage{framed}
\usepackage[nolist]{acronym}
\usepackage{microtype} 
\usepackage[ruled]{algorithm2e}

\usepackage{soul}
\usepackage{makecell}
\usepackage{pifont}
\newcommand{\cmark}{\ding{51}}%
\newcommand{\xmark}{\ding{55}}%
\DeclareCaptionLabelFormat{andtable}{#1~#2  \&  \tablename~\thetable}
    \usepackage{multicol}
    \usepackage{stackengine}
    \usepackage{etoolbox}
\newcommand{\mc}[1]{\multicolumn{1}{c}{#1}} 
\newcommand{\mybf}{\fontseries{b}\selectfont} 
\robustify\mybf
\newcommand{\hey}[1]{\textcolor{red}{{ #1}}}

\usepackage[draft]{todonotes}   

\newtheorem{definition}{Definition}
\newtheorem{problem}{Problem}

\usepackage{array}
\newcolumntype{P}[1]{>{\centering\arraybackslash}p{#1}}

\SetKwInput{KwInput}{Input}
\SetKwInput{KwOutput}{Output}

\SetCommentSty{mycommfont}

\newenvironment{captiontext}{%
   \begin{center}%
     \begin{minipage}{0.9\linewidth}%
         \sffamily\small}%
   {%
      \end{minipage}%
        \end{center}}
        
\usepackage{ragged2e}
\usepackage{adjustbox}

\newcommand{\name}{{\em NeckSense\ }}
\newcommand{\names}{{\em NeckSense}}
\newcommand{\namef}{{\em NeckSense's\ }}

\newcommand{\nameG}{{necklace\ }}
\newcommand{\nameGs}{{necklace}}
\newcommand{\nameGf}{{necklace's\ }}

\newcommand{\studya}{Exploratory Study\ }
\newcommand{\studyb}{Free-Living Study\ }
\newcommand{\studyas}{Exploratory Study}
\newcommand{\studybs}{Free-Living Study}
\newcommand{\evaluationSecond}{per-second level\ }
\newcommand{\evaluationEpisode}{per-episode level\ }
\newcommand{\evaluationSeconds}{per-second level}
\newcommand{\evaluationEpisodes}{per-episode level}

\newcommand{\noind}[0]{\vspace{5 pt} \noindent}
\newcommand{\noindpar}[1]{\noind {\bf {\small #1}}}
\setcopyright{none}

\begin{acronym}
\acro{IMU}{Inertial Measurement Unit}
\acro{LFA}{Lean Forward Angle}
\acro{GBM}{Gradient Boosting Model}
\acro{LOSOCV}{Leave-One-Subject-Out Cross Validation}
\acro{BLE}{Bluetooth Low Energy}
\acro{EMG}{electromyography}
\end{acronym}

\definecolor{orangeCol}{HTML}{ffdfa0}

\begin{document}
\setstcolor{blue}
\title[\name]{\names: A Multi-Sensor Necklace for Detecting Eating Activities in Free-Living Conditions}

\author{Shibo Zhang}
\affiliation{%
  \institution{Northwestern University}
  \city{Evanston, IL}
  \country{United States}
}
\author{Yuqi Zhao}
\affiliation{%
  \institution{Northwestern University}
  \city{Evanston, IL}
  \country{United States}
}
\author{Dzung Tri Nguyen}
\affiliation{%
  \institution{Northwestern University}
  \city{Chicago, IL}
  \country{United States}
}
\author{Runsheng Xu}
\affiliation{%
  \institution{Northwestern University}
  \city{Evanston, IL}
  \country{United States}
}
\author{Sougata Sen}
\affiliation{%
  \institution{Northwestern University}
  \city{Evanston, IL}
  \country{United States}
}
\author{Josiah Hester}
\affiliation{%
  \institution{Northwestern University}
  \city{Chicago, IL}
  \country{United States}
}
\author{Nabil Alshurafa}
\affiliation{%
  \institution{Northwestern University}
  \city{680 N Lake Shore Dr, Chicago, IL, 60611}
  \country{United States}
  \address{nabil@northwestern.edu}
}

\renewcommand\shortauthors{Zhang et al.}


\begin{teaserfigure}
  \centering
  \includegraphics[height=2.2in]{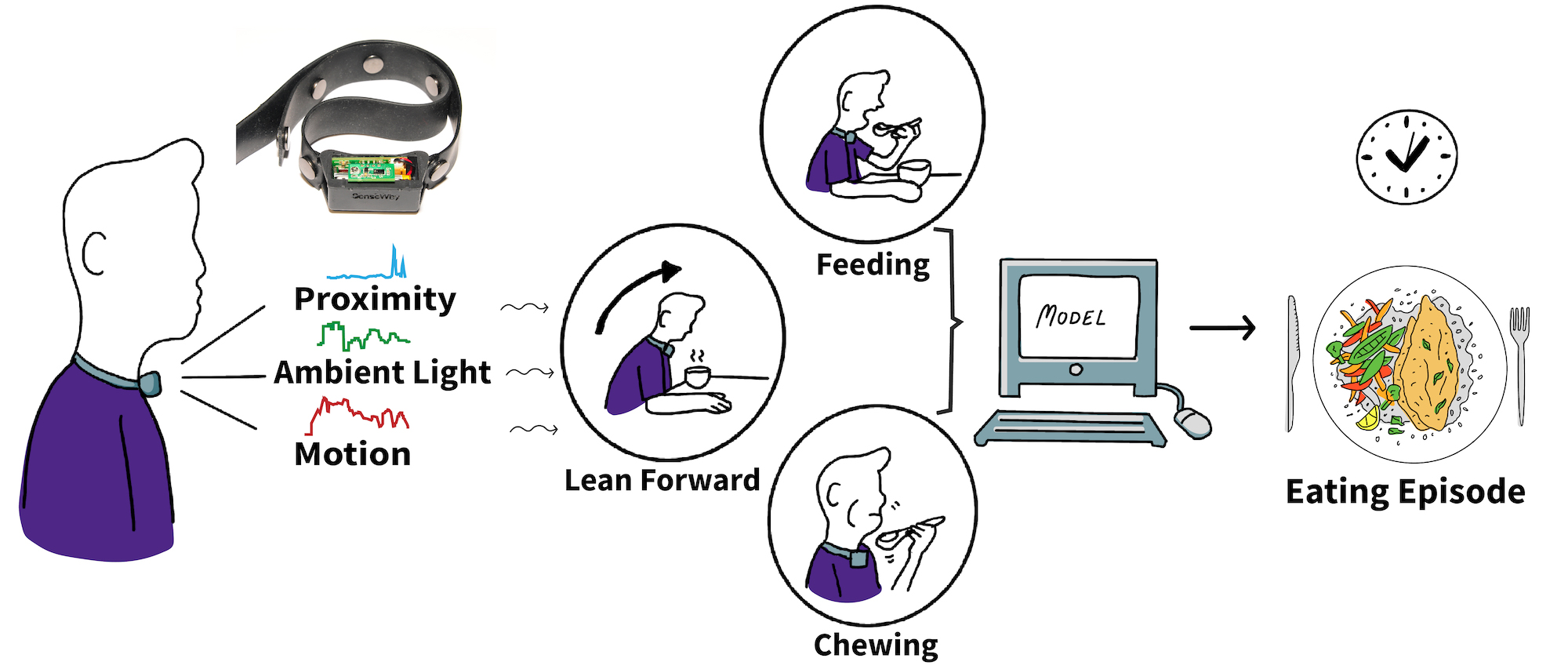}
  \caption{Our non-contact,  day-long battery life necklace~-- \names, collects proximity, ambient light and motion signals to detect chewing actions, feeding gestures, and lean forward motion that allows detection of eating episodes occurring throughout the day. \name enables long term studies for monitoring eating behavior in free-living conditions.\label{fig:introFig}\vspace{0.1in}}
\end{teaserfigure}

\begin{abstract}
We present the design, implementation, and evaluation of a 
multi-sensor low-power \nameG -- \names -- for automatically and unobtrusively capturing fine-grained information about an individual's eating activity and eating episodes, across an entire waking-day in a naturalistic setting. 
The \name fuses and classifies the proximity of the \nameG from the chin, the ambient light, the Lean Forward Angle, and the energy signals to determine \textit{chewing sequences}, a building block of the eating activity. 
It then clusters the identified chewing sequences to determine  eating episodes. 
We tested \name with 11 obese and 9 non-obese participants across two studies, where we collected more than 470 hours of data in naturalistic setting. 
Our result demonstrates that \name enables reliable eating-detection for an entire waking-day, even in free-living environments. 
Overall, our system achieves an F1-score of 81.6\% in detecting eating episodes in an exploratory study. 
Moreover, our system can achieve a F1-score of 77.1\%  for episodes even in an all-day-around free-living setting. 
With more than 15.8 hours of battery-life \name will allow researchers and dietitians to better understand natural chewing and eating behaviors, and also enable real-time interventions.

\end{abstract}



\maketitle

\input{1_intro}

\input{3_system}

\input{5_framework}

\input{4_study}
\input{6_results}

\input{2_related}
\input{8_conclusions}

\bibliographystyle{ACM-Reference-Format}
\bibliography{necklace.bib}

\end{document}

%% file: 1_intro.tex
 \section{Introduction}
\label{sec:intro}
Automatically and unobtrusively monitoring an individual's eating activity in free-living setting has been a long standing objective of the research community.
The possibility of monitoring eating activity automatically will allow individuals, researchers and ultimately clinicians to support various wellness goals in the form of interventions. For example, clinicians can design a system to trigger real-time interventions when people spend too much time eating~\cite{spruijt2014dynamic}, or a researcher can request for timely information about energy consumption in an individuals diet~\cite{doi:energyimbalance}. 
It has been well established that such interventions can help treat eating-disorders in the long-term and improve the quality of life~\cite{doak2006prevention}. 
However, it is difficult to provide such eating-related interventions without automatically detecting the eating activity itself. 
Thus, it is extremely important for any eating activity monitoring system to detect eating activity that occurs with diverse food choices, at varying times and context during the waking-day, and in myriad environments. 

A large body of work around automatic dietary monitoring has recently emerged, with wearable sensors showing promise in automatically identifying eating behaviors~\cite{illner2012review}.
Researchers have explored the possibility of detecting eating activity using wrist-based inertial sensor data~\cite{Thomaz2015,Dong2014, HASCA2017whenFail}, on-ear or on-throat based audio sensor data~\cite{amft2009overview,bi:ubicomp18,rahman2014bodybeat}, image-based information~\cite{ThomazImage2013, Reddy2007DietSense,o2013using}, or a combination of one or more of these techniques~\cite{liu2012intelligent, mirtchouk2017ubicomp, sen2015case}. 
Although several automatic eating activity monitoring techniques exist, these techniques are either obtrusive, or they have not been tested in a completely free-living setting with a diverse population, or they have not been tested in long-term studies. 

To bridge this gap, in this paper, we present the design and evaluation of \names, a \textit{multi-sensor \nameG} for automatically detecting the eating activity.
The goal of \name is to ensure that it can \textit{accurately}, \textit{automatically} and \textit{unobtrusively} monitor its wearer's eating activities that occur \textit{at any time} during the day, \textit{in any setting}, while ensuring that the device has at least an entire waking-day of \textit{battery-life}. We also want to ensure that \name generalizes and detects eating accurately for a demographically diverse population, including people with and without obesity. 

In this work, we assume that the combination of an individual's leaning-forward action, performing the feeding gesture, and then periodically chewing together constitutes the in-moment eating activity. 
We design and develop \names, a \nameG with an embedded proximity sensor, an ambient light sensor, and an \ac{IMU} sensor that can capture these aforementioned actions. 
To detect this eating activity, \name fuses and classifies features extracted from these sensors. 
It then clusters the predicted eating activity to determine an eating episode. Figure~\ref{fig:introFig} provides an overview of \names. 
We  evaluate the feasibility of \name by conducting two user studies: a longer,  intermittently-monitored free-living \textit{\studya} (semi-free-living) and another completely \textit{\studyb}. The exploratory study allowed us to identify sensors that were useful in detecting the eating activity. It also allowed us to identify usability concerns with the \nameGs. The findings of the \studya allowed us to improve \names. We evaluated the improved \nameG on a diverse non-student population for two full-days in a completely free-living scenario, while participants carried out their everyday activities. 

While designing, developing and evaluating \name to address practical challenges in free-living studies pertaining to (a) accurately monitoring the eating activity that occurs in diverse settings, (b) usability and comfort of wearing the device, and (c) collecting the sensor data energy efficiently, we make the following \textit{key contributions}: 
\begin{enumerate}
    \item We describe the design and implementation of a multi-sensor necklace for detecting eating activity and eating episodes. The necklace utilizes sensor data from its proximity, ambient light, and \ac{IMU} sensors to determine eating activity.
    \item We evaluate the necklace in {two studies: an exploratory semi-free-living and another completely free-living} to determine the possibility of detecting eating activities in naturalistic settings. The studies involved participants with varied body masses. The participants consumed 117 meals during this study period. Overall, we found that in a semi-free-living setting, the necklace could identify eating activity at an ambitious fine-grained \evaluationSecond with an average F1-score of 76.2\% and at a coarse-grained \evaluationEpisode of 81.6\%. As expected, the fine-grained performance drops to 73.7\%, and the coarse-grained performance drops to 77.1\% in a completely free-living setting.
    \item We evaluate the energy performance of the system during these studies and observed that on average the battery life of the device during the \studya was 13 hours, while the battery life improved to 15.8 hours in the \studybs with extra BLE time synchronization function.
    \item We will anonymize and make both our datasets available for use by the community. 
    The dataset contains the sensor traces collected from 20 participants, tagged with ground truth labels generated from video and clinical standard labeling practices. 
\end{enumerate}

Overall, the necklace provides a practical solution for automatically monitoring eating activity in completely free-living settings. This will enable researchers or clinicians to provide appropriate interventions to the device's user. In the near future we anticipate that other researchers will use our dataset to validate their own methods for chewing sequence, and eating episode detection. 


%% file: 3_system.tex
\section{System Design \& Implementation}
\label{sec:sys}

\begin{figure}
\centering
\begin{minipage}{.46\textwidth}
  \centering
  \vspace{0.5in}
  \includegraphics[width=\columnwidth]{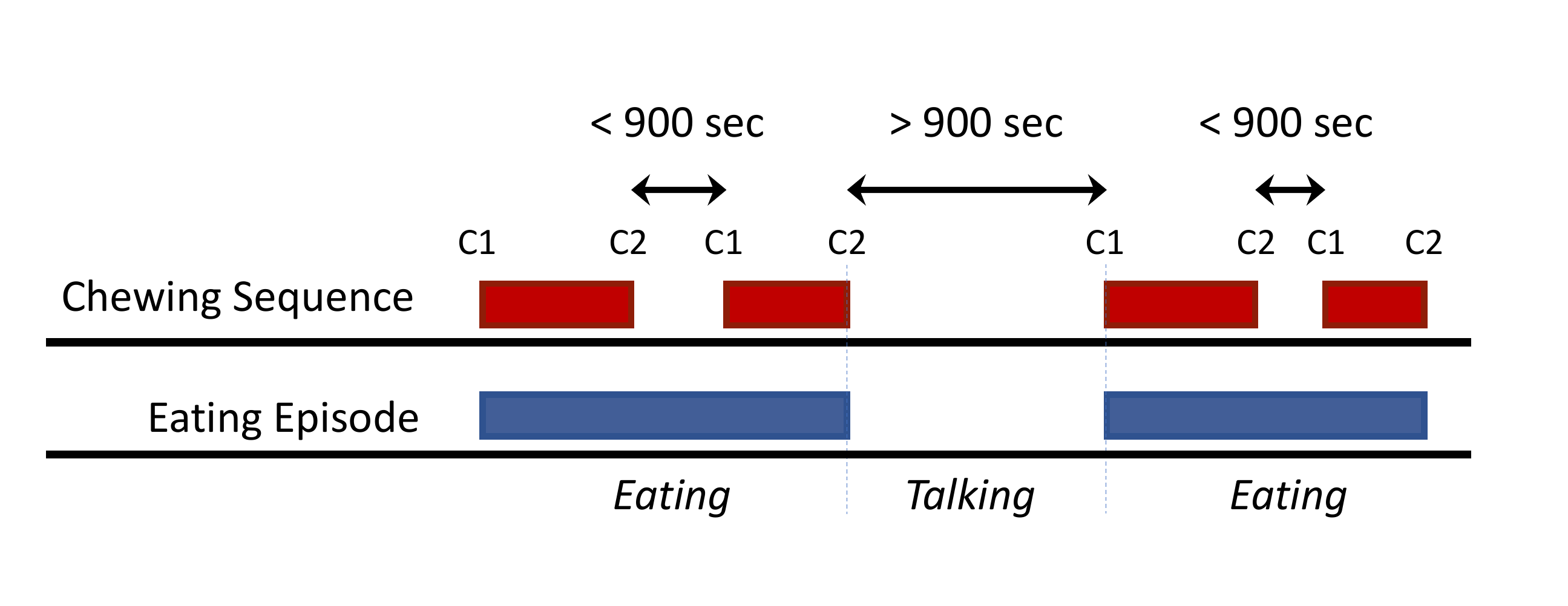}
\caption{Schematic of eating episodes that is composed of multiple chewing sequences. C1 and C2 correspond to start and end of each chewing sequences. We use an data-driven approach to determine minimum interval between chewing sequences to identify episode boundaries.} \label{fig:definitions}
\end{minipage}\hfill
\begin{minipage}{.46\textwidth}
  \centering
  \includegraphics[width=0.7\columnwidth]{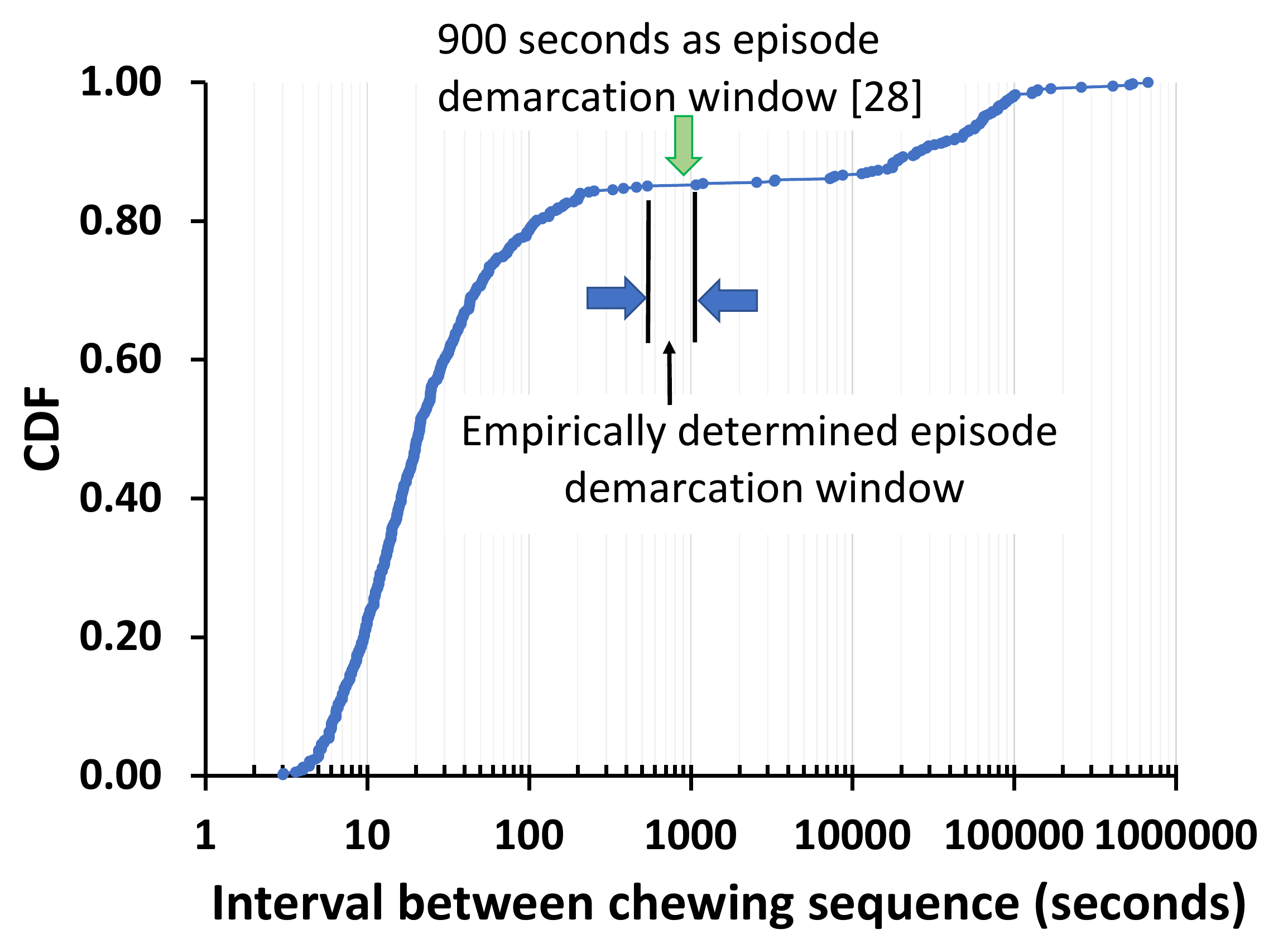}
    \caption{Cumulative distribution function representing an empirical approach to determine eating episode boundaries based on the time between the end of one chewing sequence and the start of the next chewing sequence. Our empirically determined inter-episode gap is similar to the inter-episode gap as suggested by Leech et al.~\cite{leech2015characterizing}. 
    \label{fig:choiceOf600Sec}}
\end{minipage}\vspace{0.1in}
\end{figure}

While keeping the challenges of free-living data collection in mind, we present our multi-sensor neck-worn device that is tolerant of varying sensor positions as captured in real world settings, comfortable to wear, semi-autonomous (only requiring users to turn on/off and charge the device at the end of the day), and validated using a custom-designed ground truth camera with full-day battery life (allowing uninterrupted ground truth monitoring). We next present the design of the multi-sensor eating detection system and also describe the camera used for ground truth validation. These devices were used in two user studies -- a semi-free living \studya and another completely \studybs. However, before getting into the system details, we provide our definition of eating. 

\subsection{Defining Eating}\label{sec:define_eating}
We consider the act of mastication (i.e., the act of periodically bringing the jaw together to grind food) as a \textit{chewing sequence}. 
In this paper, we define an \textit{eating episode} as an aggregate of \textit{chewing sequences} that occur within a short duration of time, and these chewing sequences are separated from other chewing sequences by a large time gap. An eating episode can represent either a snack or a meal. Figure~\ref{fig:definitions} pictorially represents a chewing sequence and an eating episode.


\textbf{Chewing Sequence:}
We define a chewing sequence as a combination of chews that occur together with breaks no greater than 3 seconds between subsequent chews. In this work we determine chews (detailed in Section~\ref{sec:framework}) by applying a prominent-peak detection algorithm on the proximity sensor data, followed by running a longest period subsequencing algorithm, and finally extracting and classifying features from proximity, \ac{IMU}, and ambient light sensors.

\textbf{Eating Episode:} 
We define an eating episode as a group of chewing sequences with inter-chewing sequence breaks no larger than $\delta$ seconds. Two adjacent chewing sequences with a gap longer than $\delta$ seconds are identified as two separate eating episodes. We use a data-driven approach to determining the value of $\delta$. Figure~\ref{fig:choiceOf600Sec} presents the CDF of the interval between subsequent chewing sequences. From the figure we can observe that a value for $\delta$ between 540 seconds and 1100 seconds provides a clear boundary between eating episodes. We decided to use $\delta = 900 $ seconds in our evaluation. This choice of $\delta$ empirically validates the inter-episode interval reported or used by researchers previously~\cite{leech2015characterizing, bi:ubicomp18}. Applying this rule allowed us to turn chewing sequence labels to eating episode labels with exemption from ambiguity when evaluating eating episode prediction.  


\begin{figure}
\centering 
\includegraphics[width=.9\textwidth]{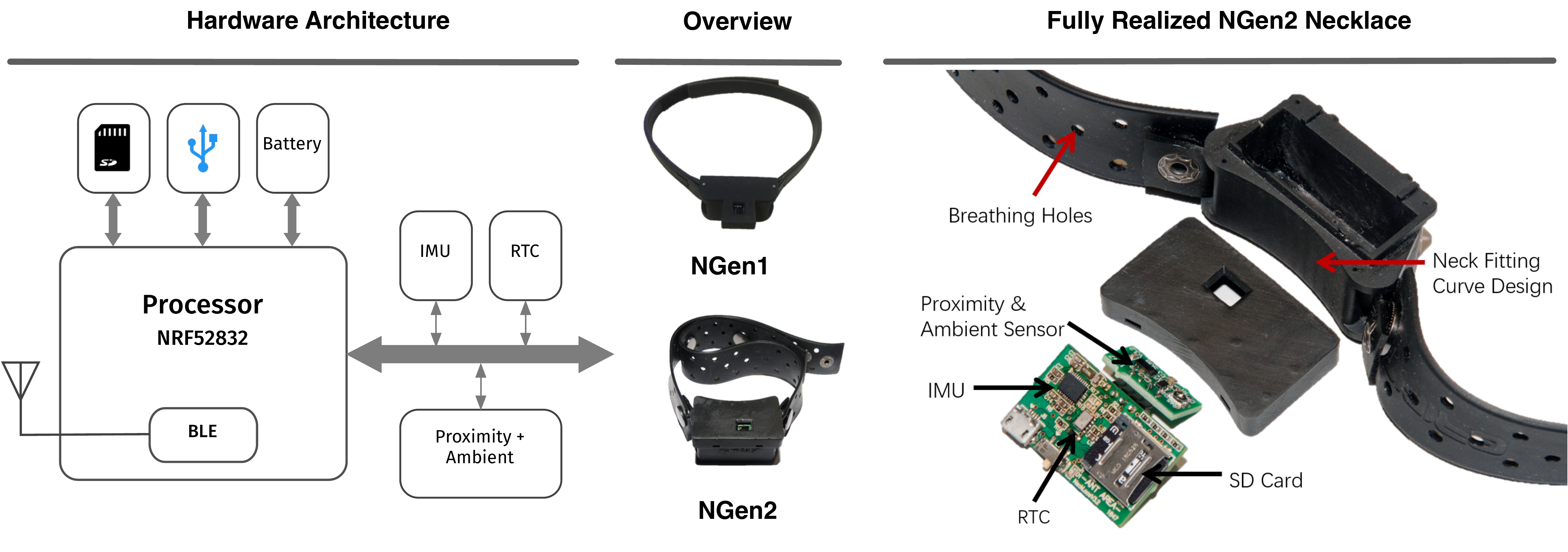}
\caption{
Hardware architecture and overview of two generations of  \nameGs. NGen1 was used in the \studyas, while NGen2 was used in \studybs. }
\label{fig:necklace-arch}
\end{figure}

\subsection{Necklace Design}
To evaluate the feasibility of using a neck-worn method for eating and chewing detection, we designed and developed a multi-sensor \nameG \textit{NGen1} that was used in the \studyas. Lessons learnt from the study helped in designing and developing the \textit{NGen2} \nameG that was used in the \studyb (details about the studies are presented in Section~\ref{sec:study}). Figure \ref{fig:necklace-arch} presents the overview and hardware architecture of the \nameGs. 

\noindpar{Signal and Sensors:} 
Our proposed system design is based on observation and validation of eating activities in laboratory settings. 
We observed numerous eating activities and noticed that during most eating activity  an individual leans forward to reach for the food; the hand feeds the food to the mouth; and the jaws continuously grind to perform the repetitive chewing action.  
To capture these eating related signatures, we evaluated several sensors and finally chose an \ac{IMU} sensor, a proximity sensor, and an ambient light sensor, all of which were embedded into a neck-worn device. 
The \ac{IMU} sensor facilitates determining \textit{leaning forward} movement, i.e., the \ac{LFA}. 
The proximity sensor on the necklace is directed towards the chin and allows monitoring the variation in signal while an individual is chewing. 
The on-necklace ambient light sensor's reading drops when the individual's hand moves towards the mouth during a feeding gesture, thus allowing detecting of the feeding gesture. 
We observed (described in Section~\ref{sec:evaluation}) that although each sensor can individually detect the eating action, fusing the signals from these sensors improves the overall detection performance. We thus use all the sensors in the final \namef design.

\noindpar{Hardware Design:} 
Both NGen1 and NGen2 are centered around a Nordic NRF52832 SiP that has an ARM Cortex M4 CPU and a \ac{BLE} communication interface. 
Two sensors chips -- VCNL4040 and BN0080 are embedded onto the necklace.
The VCNL4040 is a combination of proximity and ambient light sensor. We intend to capture the chewing action using this sensor and thus it faces upwards, towards the user's chin. 
The BN0080 is an \ac{IMU} sensor that is utilized for computing the \ac{LFA}. 
All sensors were sampled at 20Hz. 
The proximity, ambient light, and \ac{IMU} sensors sequentially write to a memory buffer. 
When the buffer reaches its capacity, the data is written to a local micro SD ($\mu$SD) card. 
The buffer helps in increasing the  writing throughput and reducing the power consumption by writing periodically. 
The whole device is powered by a 350mAh Li-on battery which guarantees more than 14 hours of battery life, making it possible to monitor eating-activities during an entire waking-day. 

\noindpar{Time keeping and Synchronization:} 
Each chewing gesture is a short-lived action and thus to ensure syncronization, it is important to maintain a highly accurate time keeping system.  
The necklace has to ensure that it always records an accurate timestamp without any manual calibration and time setting. 
NGen1 did not have an RTC module with a backup battery, and so when the main battery failed, the data had invalid timestamps, making synchronization of the necklace and camera data challenging. 
As an improvement, 
 we introduce a real-time clock (RTC) in the NGen2 necklace so that it maintains time at a 2-ppm accuracy. 
With its rechargeable backup battery, it can keep time for more than 1 year after the main power source is dead. 
The accurate RTC timestamp ensures that the system's clock does not drift substantially.  Additionally, whenever the necklace connects with a hub device (e.g., a smartphone), it automatically fetches the current time from a time server by calling the Current Time Service via BLE, the service help to correct the time and keep it during long period of study. 
We empirically observed that this feature eliminates the inevitable time drift of <180msec per day. 

\noindpar{Mechanical Design:} 
We designed the necklace along with a case and a strap, as shown in Figure~\ref{fig:necklace-arch}. 
This design has evolved over multiple iterations (with feedback from the \studyas) to ensure that in addition to being functional, it is comfortable to wear. 
The necklace's band is a sports-style porous silicone band that wraps around the neck. This was an improvement in NGen2 over NGen1 where we used a leather band. 
The case housing the PCB and sensors is connected to this band. 
The silicone material and carefully balanced weight of the case ensures that the sensor's position is stationary, even during vigorous activities. 
This property guarantees that we can collect quality data, even in noisy free-living environments. 
The resin material used for manufacturing the case (commonly used in wearable jewelry) is smooth and skin friendly. 
One face of the case in NGen2 was curved to ensure that it was comfortable on the neck (this was an improvement based on feedback from NGen1).  
Based on user-feedback, we also add breathable holes to the band to increase user comfort. 
Recent research indicates that a participant will adhere to wearing a device if the participant can decide on the device from a collection of devices with varied designs~\cite{CHI2017WillSenseRawan}. 
We thus manufactured devices with diverse designs and colors.
We anecdotally observed that when participants chose their own device, they wore it for longer duration. 

\begin{figure}[t]
\centering
\includegraphics[width=0.85\columnwidth]{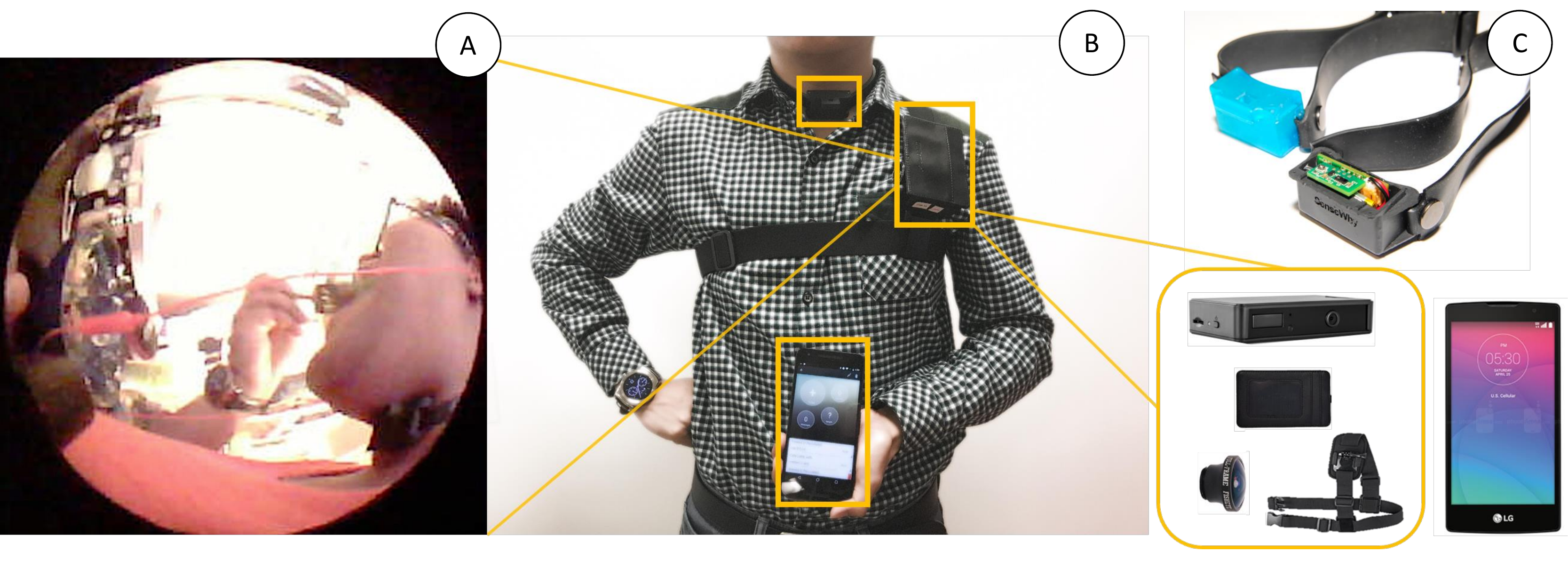}
\caption{
Participant wearing the devices during the \studyb (A): Representative still image of the ground-truth camera output showing the chin, hand, food, and utensil. (B): Participant wearing the necklace and Ground-Truth camera. (C): All devices for \studyb. 
\label{fig:validationsystem}}

\end{figure}

\subsection{Ground Truth Video Camera}
We used a fish-eye lens fitted wearable camera (that captured both audio and video) to continuously record participant behavior including fine-grained eating related activities. 
This recording provided the ground truth labels for learning models and validating free-living studies~\cite{rawanubicomp18}. 
Once the neck-worn sensor is validated, we envision that the necklace's user will not use this camera. 
Figure~\ref{fig:validationsystem} illustrates the various components of the wearable camera system. The system comprises of a Qcam QSD-722 camera, a fish-eye lens, a shoulder strap, and an ID card to conceal the device.
The camera is positioned around the shoulder with the lens facing the participant's dominant hand rather than pointing outward (e.g., as in SenseCam~\cite{hodges2006sensecam}). 
This minimizes privacy concerns of bystanders, although bystanders to the side of the participant are noticeable. At this angle, the wearable camera can capture the mouth, the jaw, the neck, and the utensil and foods consumed (see Fig.~\ref{fig:validationsystem}A).
As privacy is a significant concern for the participant, we provided participants with the option of deleting segments of video data they did not want study supervisors to see.

%% file: 5_framework.tex

\section{Eating Detection Framework}
\label{sec:framework}
\begin{figure}
\centering
\begin{minipage}{.46\textwidth}
  \centering
  \includegraphics[width=0.9\columnwidth]{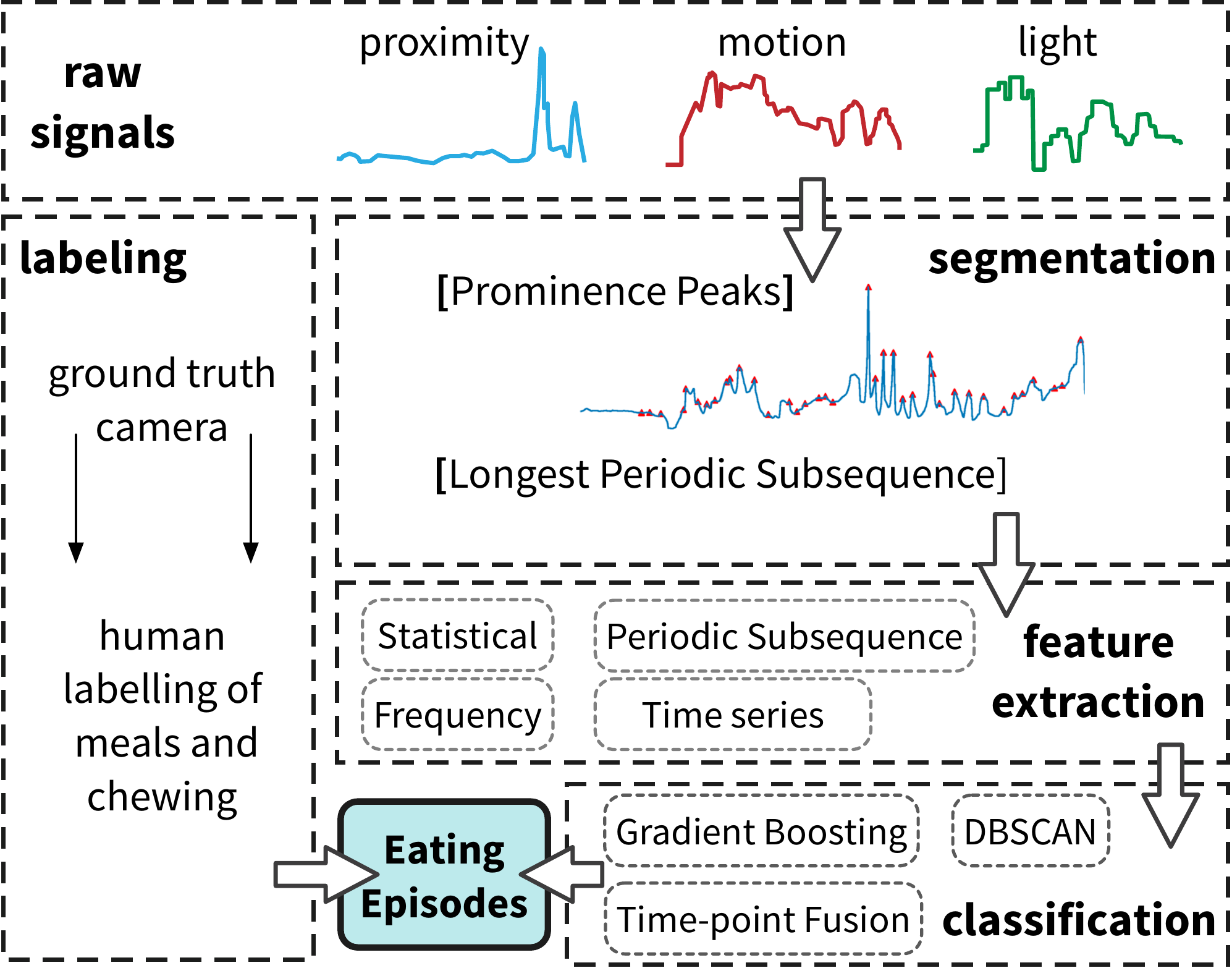}
\caption{Chewing and meal detection framework and validating it with the labeled ground truth.}
\label{fig:framework-overview}
\end{minipage}\hfill
\begin{minipage}{.46\textwidth}
  \centering
  \includegraphics[width=0.9\columnwidth]{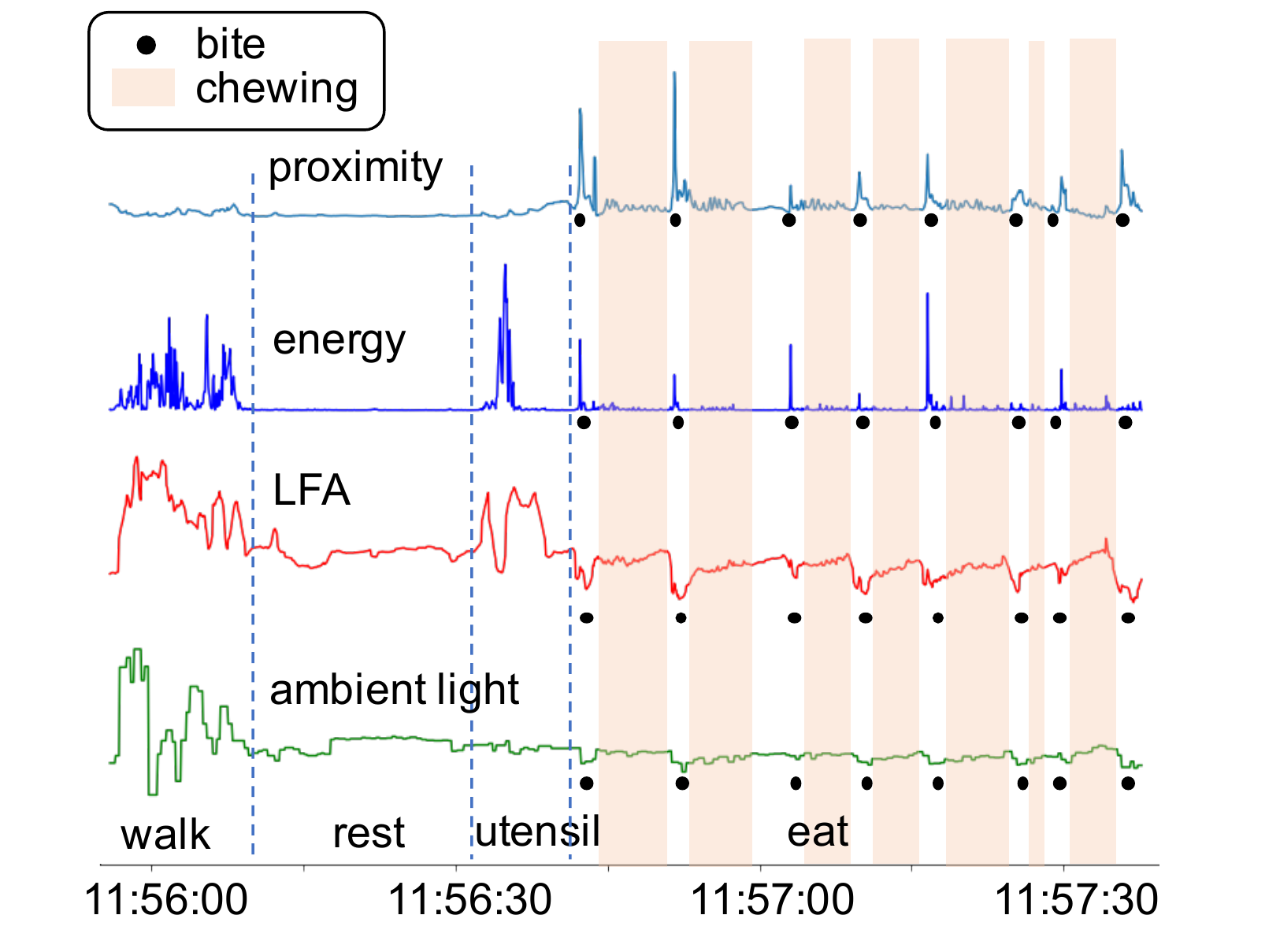}
\caption{Four signals (proximity, energy, LFA, and ambient light) captured while the wearer is walking, resting, utensiling, and eating.}
\label{fig:signals-v-activity}
\end{minipage}
\vspace{0.2in}
\end{figure}

We next present our method of predicting the chewing sequences and eating episodes from the signals generated by the sensors. 
The entire processing pipeline is presented in Figure~\ref{fig:framework-overview}. 
The pipeline consists of five steps that we shall next describe. 
\subsection{Signal pre-processing}
The signals extracted or calculated from the necklace's sensors includes (i) proximity to chin, (ii) ambient light value, (iii) \ac{LFA}, and (iv) energy signal defined as the sum of squares of tri-axial accelerometer values.

(i) Proximity to chin: We extract the proximity to the chin from the \nameGf proximity sensor. The sensitivity range of the proximity sensor is set between 5 and 250 mm. Since the average neck length of a human being varies between 100 and 120 mm, this setting should sufficiently capture the chewing action.  
Figure~\ref{fig:signals-v-activity} shows representative signals when the participant was walking, resting, utensiling (using utensils, but not eating), and eating. During the eating activity, smaller peaks in the proximity sensor's signal signifies chews, while the larger peaks signify bites.

(ii) Ambient light value: The \nameGf ambient light sensor provides the ambient light value. 
The ambient light value is highest when the user turns their head right or left, since the sensor is not obstructed by the head. 
The value is lowest when the user lowers their head and move their hand towards their mouth during a feeding gesture. We can see the periodic drop in ambient light values during the eating activity in Figure~\ref{fig:signals-v-activity}.

(iii) \ac{LFA}: We obtained the necklace's absolute orientation in the form of quaternions from the on-necklace \ac{IMU} sensor. 
The quaternion is a 4D vector $\bm{q}$ representing the rotation axis and the angle of rotation around that axis. 
$\bm{q}$ can be projected into different planes to gain physical angles and infer activities such as leaning forward and to the side, and to determine the orientation of the wearer. 
However, not all of these angles are related to the eating process. 
The most informative angle is the \ac{LFA}, the angle between the \ac{IMU} and earth's surface. When the wearer sits straight, the \ac{LFA} is close to 90$^\circ$.
\ac{LFA} is calculated by applying the dot product of the normal vectors of two planes:

\begin{center}
$LFA = acos <\bm{n_1}, \bm{n_2}>$
\end{center}

where the normal vector of Earth's surface is the z-axis, and the normal vector of the \ac{IMU} is obtained through the quaternion transformation:
\begin{center}
  $\bm{n_1} = [0, 0, 1]$
\qquad
\qquad
$\bm{n_2} = \bm{q}\bm{n_1}\bm{q^{-1}}$
\end{center}

where $q$ is a unit quaternion that rotates $n_1$ to obtain the normal vector of the IMU. It is worthwhile to note that while LFA does not always occur in the field, particularly when snacking while sitting on a couch, features from \ac{LFA} can enhance detection of bites. 

(iv) Accelerometer's energy value: The on-necklace \ac{IMU} also provides the tri-axial accelerometer data  $(a_x, a_y, a_z)$ capturing acceleration from the three axes. 
We calculate the energy signal as the sum of squares of the tri-axial acceleration components, $E=a_x^2+a_y^2+a_z^2$. 
Features computed from $E$ help reduce false positive rates generated during physical activities.
From Figure~\ref{fig:signals-v-activity} we can see that during the eating gesture there are peaks in the energy signal. However, unlike the peaks observed in the signal during the walking activity, the peaks observed during eating are sparse. 


\subsection{Labeling}
The process of labeling the data allows establishing the ground truth.  
Annotators labeled the start and end of every chewing sequence by visually and acoustically confirming the information using the video captured by the camera.  
If a participant continuously chewed during her entire meal, the entire episode was labeled as one chewing sequence. Annotators marked the end of a chewing sequence when the participant stopped chewing for at least 3 seconds. For example, if the participant took four breaks that were each at least 3 seconds long during the meal, there were five chewing sequence labels in that meal.

From the labeled chewing sequences, the annotators identified the eating episodes. As discussed in Section~\ref{sec:define_eating}, we define an eating episode as a group of chewing sequences with chewing sequence breaks no longer than 900 seconds. 
Any two adjacent chewing sequences with a gap longer than 900 seconds were regarded as two separate eating episodes. 
Applying this rule allowed the annotators to establish eating episode labels from the chewing sequence labels.

\subsection{Segmentation}
\label{sec:framework_segmentation}

We employed time-based methods on the necklace's proximity signals to detect periodic chewing patterns. 
To identify candidate chewing sequences, we applied a peak finding algorithm followed by a periodic subsequence algorithm that detected candidate chewing subsequences.

\begin{figure}[t]
\centering
\includegraphics[width=0.45\columnwidth]{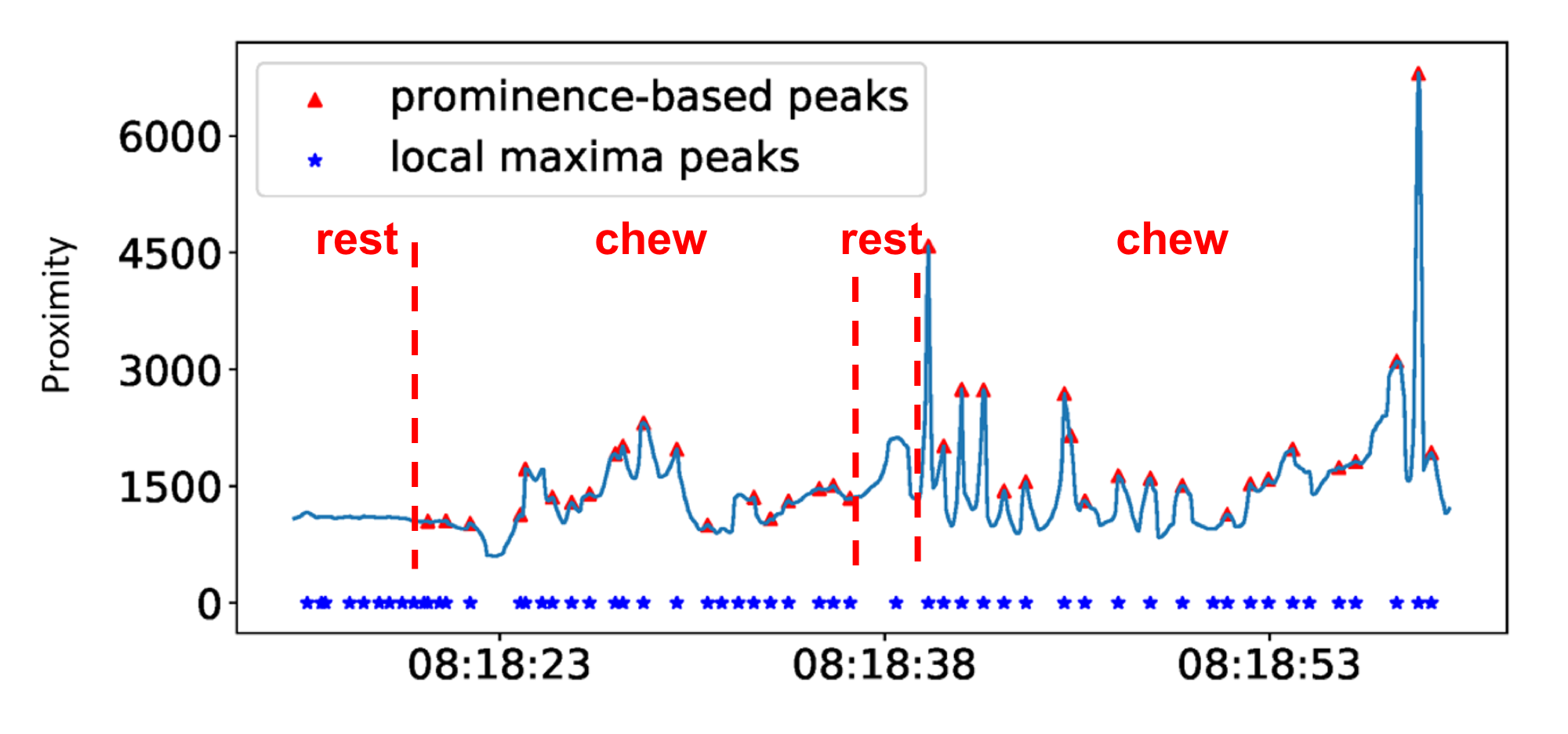}
\caption{Prominence-based peak finding algorithm (height=4.5) vs local maxima peaks (using 2 samples before and after the time point.)}
\label{fig:peaks}
\end{figure}

\subsubsection{Prominence-based peak-finding algorithm}
The first step in detecting periodic sequences is to identify peaks in the proximity signal. 
We applied a typical peak finding algorithm that returned both prominent and non-prominent peaks (non-prominent peaks may be associated with noise and other activities). Chewing peaks are often prominent peaks that stand out due to their intrinsic height and location relative to other nearby peaks. For example, in Figure \ref{fig:peaks} we can observe several local maximum, yet non-prominent peaks that are identified during the resting period. However, more prominent peaks are apparent during the eating (chewing action) period.

\subsubsection{Longest Periodic Subsequence Algorithm}
We adapted the longest periodic subsequence algorithm to identify chewing peaks that were $\epsilon$-periodic~\cite{gfeller2011periodic}. The time points of the peaks from the prominence algorithm generated a sequence of timestamps for each peak. In this section we explain the significance of $\epsilon$-periodic, define the periodic subsequence problem, and present a dynamic programming solution for the problem.

\begin{definition}
\textbf{$\epsilon$-periodic}: 
Given a sequence of increasing timestamps $t_i$, where $i \in \{1 \ldots N\}$, the difference between consecutive numbers is $p_i = t_{i+1} - t_{i}, \forall  i=$\{$1 \ldots (N-1)$\}.  If $p_{min}$ and $p_{max}$ are the smallest and largest values of these differences, respectively, then the sequence is defined to be $\epsilon$-periodic if:

\begin{center}
$\frac{p_{max}}{p_{min}} < 1+\epsilon$
\end{center}

\end{definition}

\begin{problem}
\textbf{Relative error periodic subsequence}: 
Given a sequence of increasing numbers $t_i$ find all longest subsequences that are $\epsilon$-periodic.
\end{problem}

\begin{problem}
\textbf{Absolute error periodic subsequence}: 
Given a sequence of increasing numbers $t_i$ find all longest subsequences such that consecutive differences are bounded by $p_{min}$ and $p_{max}$.
\end{problem}

Problem 1 is not trivial when the lower $p_{min}$ and upper bound $p_{max}$ are not known. 
However, given that the chewing frequency range is known~\cite{po2011chewing}, these bounds can be estimated. 
The problem can then be solved by evoking multiple calls to a function that implements 
the absolute error periodic subsequence problem. 
Each time a function call is made by passing a new $p_{min}$ and $p_{max}$ that starts from the smallest inter-chew distance $min$, all the way until the largest inter-chew distance $max$, incrementing $p_{min}$ by multiples of $(1+\epsilon)$. 
Chewing activity has been shown to mainly occur in the range of 0.94 Hz ($5^{th}$ percentile) and 2.17 Hz ($95^{th}$ percentile), as a result we set $min=0.4$ seconds, and $max=1.5$ seconds. 
We solve the absolute error periodic subsequence problem using dynamic programming, by defining the following recurrence:

$$OPT[i] = \max_L{OPT[j]}, \ \forall j \  \ni \  p_{min} < t_i-t_j < p_{max}$$

\begin{figure}
\centering
\begin{minipage}{.46\textwidth}
  \centering
  \includegraphics[ height=2.25in]{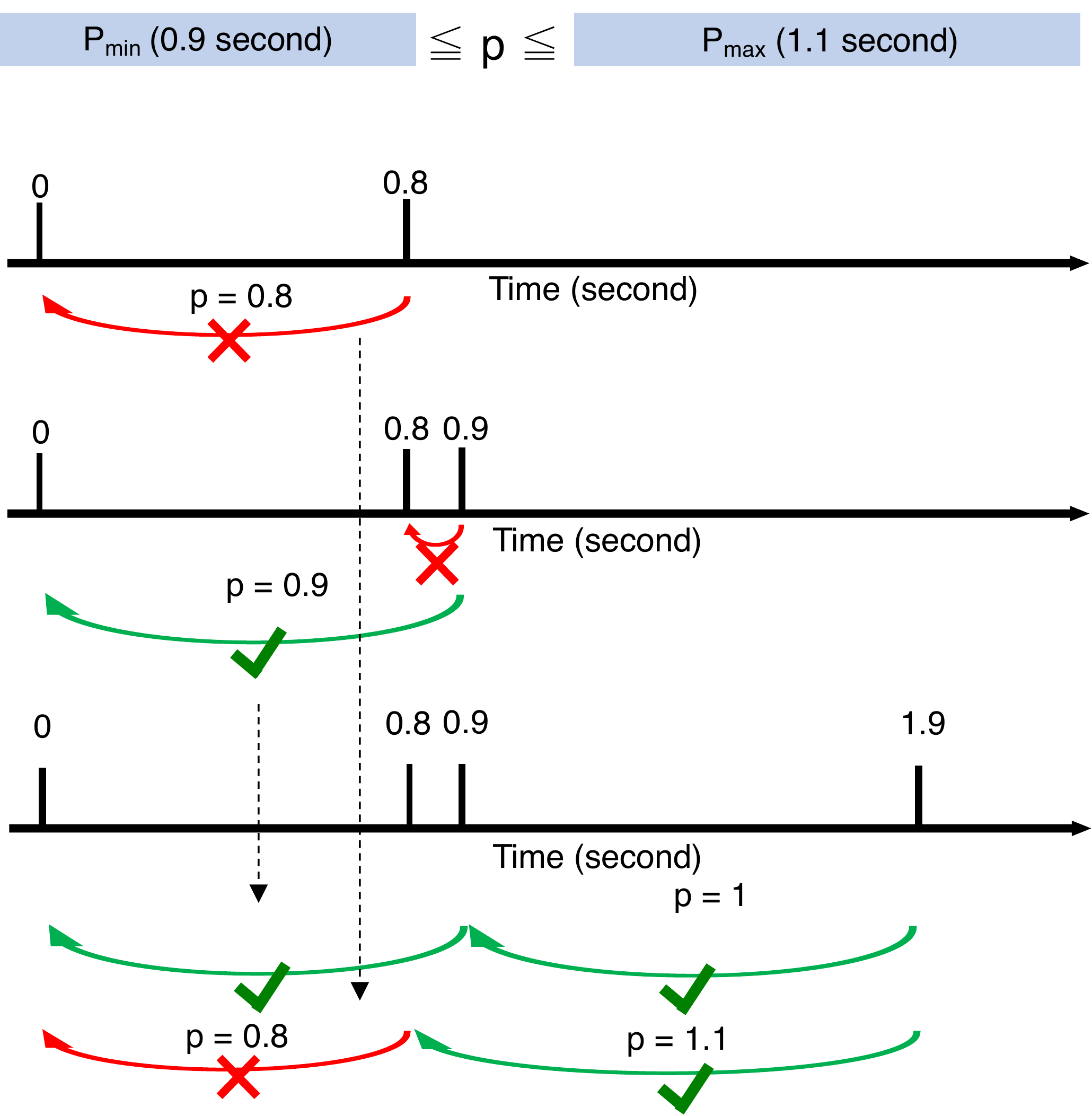}
\caption{Dynamic programming solution for absolute error periodic subsequence of proximity signal.}
\label{fig:dp}
\end{minipage}\hfill
\begin{minipage}{.46\textwidth}
  \centering
  \includegraphics[width=\linewidth]{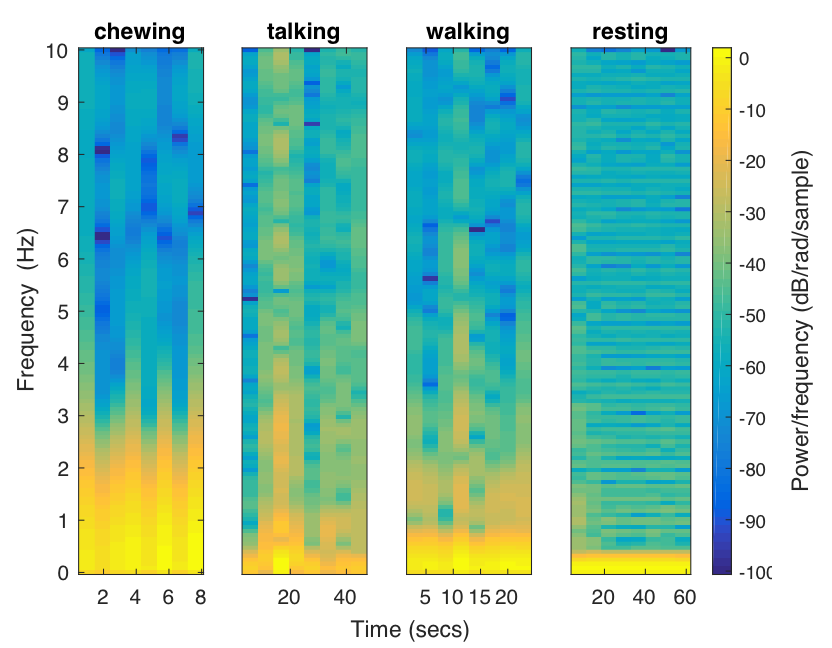}
\caption{Spectrogram of proximity signal for chewing, talking, walking, and resting.} 
\label{fig:spectrogram}
\end{minipage}
\vspace{0.2in}
\end{figure}

The absolute error periodic subsequence algorithm is called by passing the entire array of timestamps $t[i]$, a value for $p_{min}$ and $p_{max}$. It then iterates through the entire array, and for every index $i$ it calculates the longest subsequence that ends at index $i$, searching only the previous indices in the array that satisfy the inter-chew distance $p_{min}$ and $p_{max}$. Figure~\ref{fig:dp} shows an example where prominence peaks are detected at times 0, 0.8, 0.9, and 1.9 seconds. If $p_{min}$ is 0.9 and $p_{max}$ is 1.1 seconds, then the optimal subsequence is of length 2, at timestamps (0,0.9,1.9) seconds. Sequence (0,0.8) is not valid because the difference between 0.8 and 0 is 0.8 which is <$p_{min}$, the same as sequence (0.8,0.9) .

The time complexity analysis of the absolute error periodic subsequence algorithm is $\mathcal{O}(N)$, assuming the valid number of predecessors is constant. If the distance between $p_{min}$ and $p_{max}$ is a function of $N$, then this assumption does not hold. However, in practice, the difference between the smallest inter-chew distance $min$, and the longest inter-chew distance $max$ is a small constant.

\subsection{Feature Extraction}\label{sec:featExtract}

We extracted features from all the identified periodic subsequences 
to classify and validate whether the candidate subsequence identified in the \textit{Segmentation  step} (Section~\ref{sec:framework_segmentation}) was truly a chewing sequence. We extracted statistical-based features which are known to be useful in detecting physical activity~\cite{Alshurafa13} and eating~\cite{Thomaz2015}, including: maximum, minimum, mean, median, variance, root mean square (RMS), correlation, skewness, kurtosis, 1st and 3rd quartile values, and inter-quartile range. We plotted the spectrogram of the proximity signal for the dominant chewing frequencies (refer to Figure~\ref{fig:spectrogram}) and observed that the dominant frequencies during chewing occurs between 1 and 2 Hz. As a comprehensive measure, we captured the amplitude of the dominant frequencies from 0.25 to 2.5 Hz. 

Given a candidate sequence with start and end time $[c_1,c_2]$, the features listed in Table~\ref{tab:featDesc} are calculated from two window lengths. The first window $CW = [c_1 - 2sec, c_2 + 2sec]$, referred to as the \textit{ChewingWindow}. The second window is $BW = [c_1 - 2sec, c_1 + 2sec]$, referred to  as the \textit{BiteWindow}. Features from both these windows were concatenated into a single feature vector. $CW$ is useful in capturing information related to the chewing segment, while $BW$ captures bite-related features that occur at the beginning and end of the chewing sequence. Overall, we extracted 257 features for every sequence.


We found that the statistical-based features of frequency, particularly kurtosis, which is a descriptor of the shape of the frequency distribution, distinguishes chewing from other confounding activities. 
Also, the asymmetry of the frequency distribution, the skewness, is a unique predictor of chewing activities. 
Since inter-chewing intervals can also affect the predictability of a subsequence, we incorporated periodic subsequence-based features such as $p_{min}$, $p_{max}$, $\epsilon$, and length of the subsequence. 
Another important feature that filtered any subsequences that occurred at odd times of eating was hour-of-day. 
Although hour-of-day is not a critical feature in laboratory studies, it is an important feature for data collected over multiple days in naturalistic settings. 

\begin{table}[t]
\centering
\footnotesize
\begin{tabular}{ ll }
\toprule
    {Category} & {Features} \\
\midrule
     Statistics & Max, min, mean, median, std. dev., RMS, correlation, skewness, kurtosis, 1st \& 3rd quartile, interquartile range\\
     \hline
     Frequency & Frequency amplitude of 0.25 Hz, 0.5 Hz, 0.75 Hz, 1 Hz, 1.25 Hz, 1.5 Hz, 1.75 Hz, 2 Hz, 2.25 Hz, 2.5 Hz\\
      \hline
    Statistics of Frequency & Skewness and kurtosis of spectrum from frequency features\\ 
\hline
Time-series & Count below/above mean, First location of min/max, longest strike below/above mean, number of peaks\\
\hline
Periodic subsequence & $p_{min}$, $p_{max}$, $\epsilon$, length\\ 
\hline
     Time &  Hour of day\\
\bottomrule
\end{tabular}
\caption{List of features extracted for classification}
\label{tab:featDesc}
\end{table}

\begin{figure}
\centering
\includegraphics[width=0.9\columnwidth]{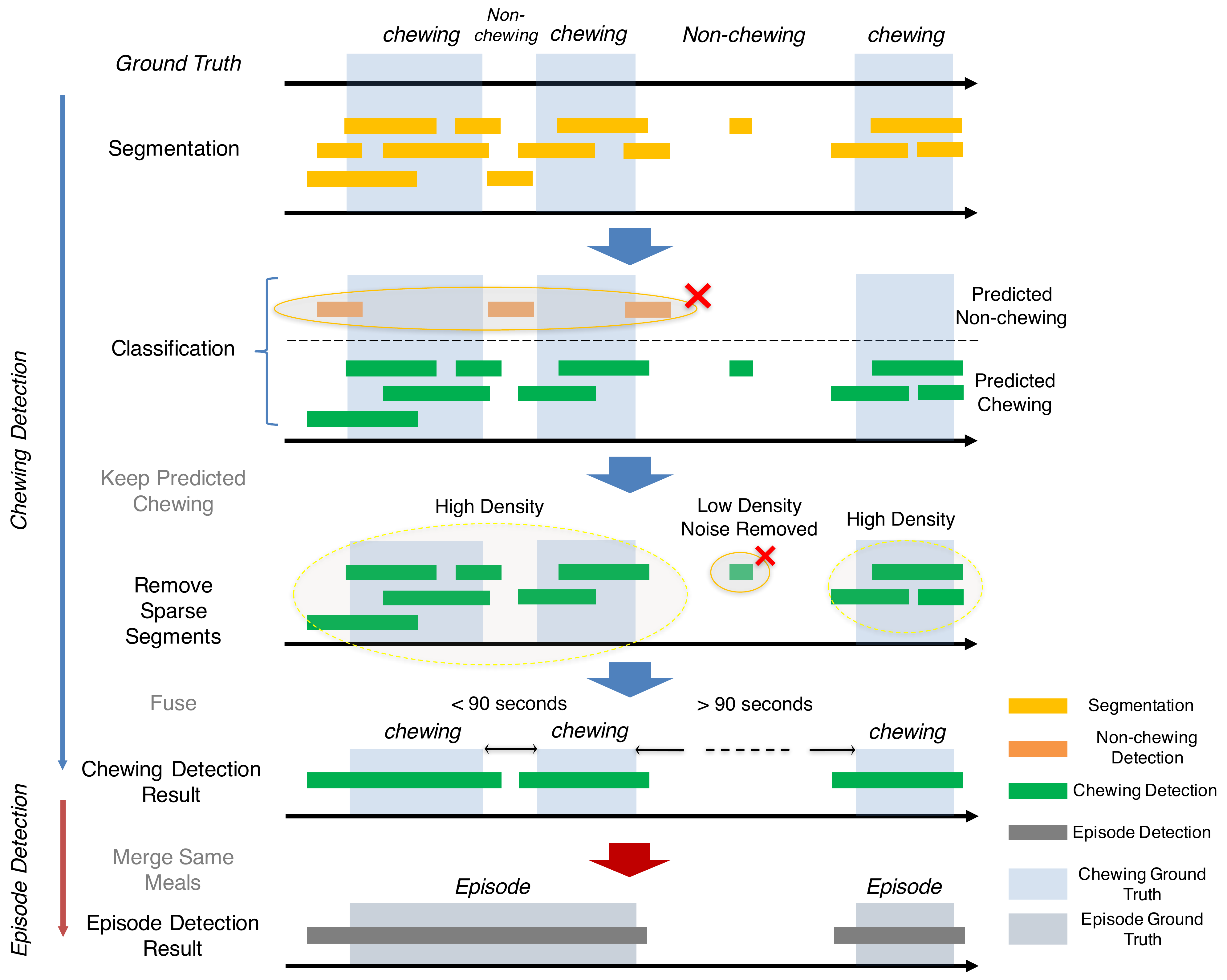}
\caption{After periodic subsequences are classified, DBSCAN filters out predicted single chewing subsequences and clusters the remainings into meals. If a ground truth eating episode has overlap with a predicted eating episode, then it is a True Positive. If a predicted eating episode has no overlap with ground truth eating episode, then it is a False Positive.}
\label{fig:eatingepisodemetric}
\end{figure}

\subsection{Classification} 
We used a gradient boosting classifier based on Friedman's \ac{GBM}, which is an ensemble classifier comprising of multiple weak learners (high bias, low variance)~\cite{GradBoost} to train a model that classifies each subsequence as belonging to a chewing sequence or not. 
We employed an open source software package, XGBoost~\cite{XGBOOST} for classification.
Gradient boosting uses regularized model formalization that controls over-fitting the training data, providing better generalizable performance.

Gradient boosting has several parameters that can be tuned to optimize its performance including general, booster, and learning task parameters. 
For the general parameters, we used the gbtree model. 
For the booster parameters, we optimized the learning rate ($eta$), the maximum depth of the a tree ($max\_depth$), the minimum loss reduction required to make a further partition on a leaf node of a the tree ($gamma$), the minimum sum of instance weight needed in a child ($min\_child\_weight$), and the subsample ratio of the training instance ($subsample$). 
We performed binary classification (chewing vs others) using a softmax objective.

Every positively detected chewing subsequence is then combined to generate a time-point distribution of predicted chewing sequences (at the per-second level). This distribution is a type of confidence weighting to estimate the likelihood of the duration belonging to a chewing sequence. Each second is then converted into a score according to the number of overlapping predicted chews. We then apply the DBSCAN algorithm 
to cluster the data. The sequences with low weights and sparse positioning are filtered during the clustering step. We then perform eating episode evaluation at the coarse-grained event level and fine-grained per-second level. Figure~\ref{fig:eatingepisodemetric} provides an overview of the processing pipeline.

%% file: 4_study.tex
\input{data/Table2_20Participants.tex}
\section{Study Design \& Data Collection} 
\label{sec:study}
Demonstrating usefulness of our eating activity monitoring system necessitates that we test the system on a representative population,  throughout the day while participants carry out their daily routine. 
Prior work has demonstrated that systems that are evaluated only in-laboratory settings often perform poorly in naturalistic settings. 
This performance degradation is quite pronounced in eating and behavior-tracking devices, as the behavior and habits of participants can easily be influenced by the in-lab setting, and the short duration of sessions rarely capture numerous real-life situations. 

With this context, we conducted an \textit{\studya} for optimizing various system parameters.  
Using the learning outcome from the \studyas, we conducted a \textit{\studyb} to determine the system's performance in a completely free-living condition. Both studies were conducted in naturalistic settings, while the participants performed their everyday activities. 
We recruited 20 participants (10 for the \studya and 10 for the \studyb) from an urban Midwestern city in the U.S. using paper flyers and via \href{https://www.researchmatch.org/}{ResearchMatch}. 
The inclusion criteria was 18-63 years of age and BMI above 18. The exclusion criteria included anyone unwilling to wear the study devices (due to a history of skin irritations or device sizing limitations) and anyone who did not own a smartphone. None of the participants were members of our research team. 
Overall, we used 134.2 hours of data from the \studya to fine-tune our system and 137.1 hours of data from the \studyb to evaluate our system.
Specific details about the data collection is presented in Table~\ref{tab:hoursinwild_demographics2}. We plan to anonymize and release this dataset for use by clinicians and researchers for evaluating their own devices and approaches to eating detection. 

During their first laboratory visit (for both \studya and \studybs), we trained the participants about how to wear and charge their devices. 
After the final day of the study, participants returned the devices and completed a post-study survey. During this visit, participants were given the option to review captured video and remove segments they felt uneasy about sharing.

\subsection{\studya}
To determine the eating sensing system's feasibility, we recruited 10 participants (4 males, 6 females, aged between 19 and 54 years) and instructed them to wear the prototype device, including the NGen1 necklace for two weeks. Participants were free to wear the device for as many or as few hours during this study. However, we instructed them to wear the prototype device during as many meals as possible.
Since we are interested in ensuring that our system performs reliably across a varied BMI range, 50\% of the participants in this study were categorized as obese (BMI>30 kg/m$^2$). The BMI of the 10 participants ranged between 21 kg/m$^2$ and 46 kg/m$^2$. 
Participants were compensated monetarily for their time (smartwatch and \$100; total value of \$300). 
Overall, we collected 277.1 hours of data during the \studyas. However, after removing data that had syncronization issues or raised privacy concerns, we acquired 134.2 hours of usable data from the \studya for our analysis.

\subsection{\studyb}
After optimizing reliability and usability of the neck-worn sensor, we designed the newer necklace device NGen2. To test this device, we recruited 10 participants (5 obese, 5 non-obese) to participate in a 2-day \studybs. 
None of these participants had participated in the previous \studyas. Participants were aged between 24 and 60 years and their BMI ranged from 20.1 to 38.1 kg/m$^2$. 
Table \ref{tab:hoursinwild_demographics2} summarizes the device usage for each participant. 
Unlike \studyas, for this study we instructed the participants to wear the device during the entire waking-day, removing only when the device was completely discharged or when they had privacy or other concerns. 
Participants did not delete any data in this study. 
Overall, after removing data segments with synchronization issues, we extracted 137.1 hours of usable data. 
We provided a \$50 compensation to the participants for their time participating in the study.

%% file: data/Table2_20Participants.tex
\begin{table}[t]
\centering
\begin{tabular*}{\textwidth}{lccccc|lccccc}
    \toprule
    \multicolumn{6}{c}{\studya} & \multicolumn{6}{c}{\studyb}\\
    \cmidrule(lr){1-6} \cmidrule(lr){7-12}
        & \multicolumn{2}{c}{Necklace}
        & \multicolumn{2}{c}{Overlap w/ Videos} 
        & &
        & \multicolumn{2}{c}{Necklace}
        & \multicolumn{2}{c}{Overlap w/ Videos} &\\
    \cmidrule(lr){2-3} \cmidrule(lr){4-5}\cmidrule(lr){8-9}\cmidrule(lr){10-11}
    \thead{Participant} & \thead{Total\\hours} &  \thead{Ave. Per \\Day}  & \thead{Total\\hours} &  \thead{Ave. Per \\Day} & Meal  & \thead{Participant} &\thead{Total\\hours} &  \thead{Ave. Per \\Day}  & \thead{Total\\hours} &  \thead{Ave. Per \\Day} & Meal  \\
    \hline
   \rowcolor[HTML]{FFDFA0} P1   & 35.7 & 5.9 & 27.8 & 5.6 & 7 &P11   & 23.9 & 11.9 & 17.2 & 8.6 & 4\\ 
   \rowcolor[HTML]{FFDFA0} P2   & 8.7 & 1.7 & 3.5 & 0.9 & 7& P12   & 14.6 & 7.3 & 7.6 & 7.6 & 1 \\ 
    \rowcolor[HTML]{FFDFA0}P3   & 74.6 & 6.8 & 31.6 & 2.9& 15 & P13   & 12.4 & 6.2 & 12.4 & 6.2 & 5\\ 
   \rowcolor[HTML]{FFDFA0} P4   & 20.3 & 3.4 & 10.1 & 3.3& 5 & P14   & 15.1 & 7.6 & 12.8 & 6.4 & 3\\ 
   \rowcolor[HTML]{FFDFA0} P5   & 40.4 & 10.1 & 33.3 & 8.3 & 11 & P15   & 19.2 & 9.6 & 8.7 & 8.7 & 4\\ 
    \cellcolor[HTML]{FFDFA0}P6   & \cellcolor[HTML]{FFDFA0}19.3 & \cellcolor[HTML]{FFDFA0}3.9 & \cellcolor[HTML]{FFDFA0}4.9 & \cellcolor[HTML]{FFDFA0}1.6& \cellcolor[HTML]{FFDFA0}9 & P16   & 20.2 & 10.1 & 16.9 & 8.5 & 5 \\ 
    P7   & 23.0 & 7.7 & 14.0 & 7.0& 3 & P17   & 23.2 & 10.6 & 15.1 & 7.6 & 6 \\
    P8   & 20.1 & 2.5 & 5.2 & 0.7& 11 & P18   & 14.9 & 7.5 & 11.0 & 5.5 & 4\\
    P9   & 20.0 & 10.0 & 0.7 & 0.4& 2 & P19   & 20.2 & 10.1 & 13.6 & 6.8& 3\\
    P10   & 15.0 & 2.5 & 3.1 & 0.8& 6 & P20   & 27.0 & 13.5 & 21.8 & 10.9 & 5 \\
    \midrule
    Total & 277.1 & - & 134.2 & - & 76 & Total & 193.0 & - & 137.1 & -& 40\\
    \bottomrule
\end{tabular*}
\caption{
Number of hours of video and necklace data in the \studya and the \studybs. Necklace valid hours exclude data with incorrect timestamp resulting from RTC going out of battery or when the necklace was not worn by the participant. Participants with obesity are highlighted in \colorbox{orangeCol}{orange}. Days are unique days in the data. Only data captured by both the camera and necklace were used in validation. From the 271.3 hours analyzed in \studyas, 14.3 hours correspond to eating, while 257 hours are non-eating.
\label{tab:hoursinwild_demographics2}}   
\end{table}

%% file: 6_results.tex
\section{Evaluation and Results}\label{sec:evaluation}
As stated in Section~\ref{sec:intro}, the overall goal of \name is to detect eating activity, while ensuring that the device has acceptable battery life. We thus evaluate \name using the \studya and \studyb data, while answering the following questions:
\begin{itemize}
    \item $\mathsf{Q1:}$ Can \name effectively detect eating activity? How does the system perform as compared to other similar techniques? 
    \item $\mathsf{Q2:}$ How do factors such as sensor choice, device position and classification features affect the detection performance? 
    \item $\mathsf{Q3:}$ Can \namef battery support monitoring an individual's eating activities throughout the waking day?
\end{itemize}

Before answering the questions, let us describe the evaluation metric used to evaluate \names.

\begin{figure}
    \centering
    \includegraphics[height=1.3in]{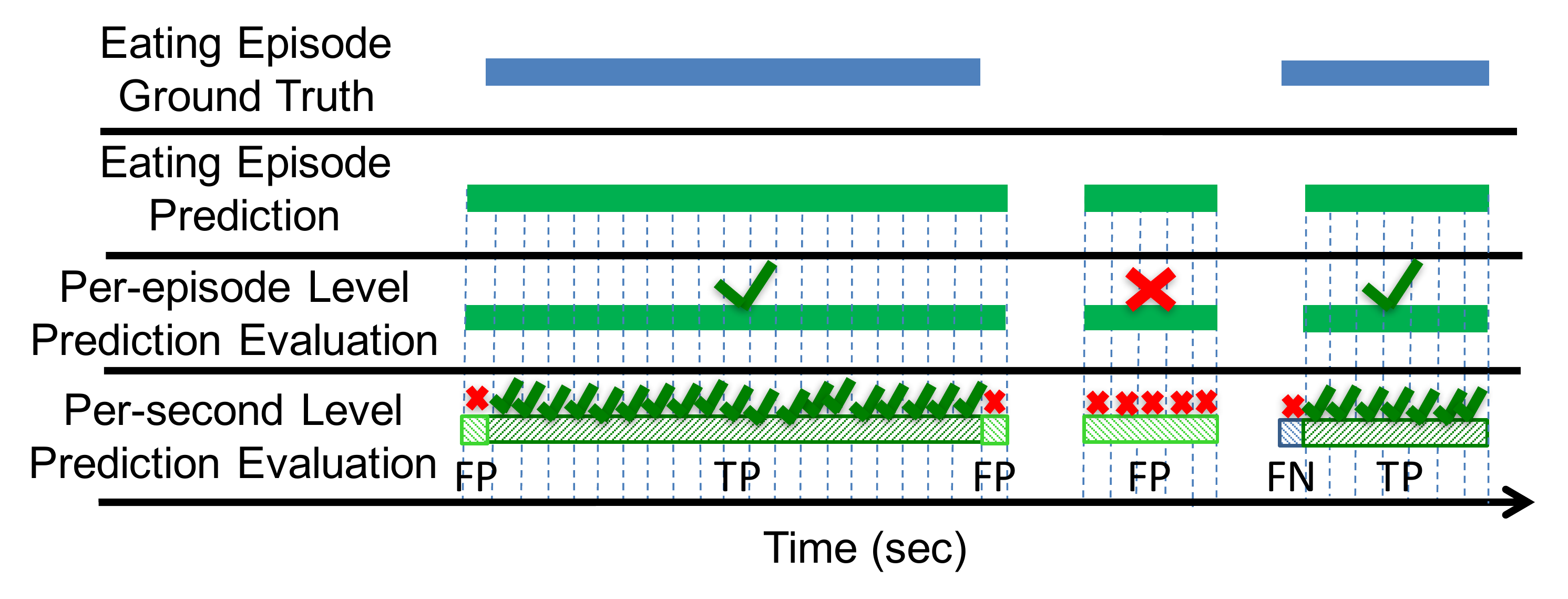}
    \caption{Evaluation criteria for the eating activity detection at two levels -- a commonly used \evaluationEpisode evaluation approach and a challenging \evaluationSecond evaluation approach. If there is a {50\%} overlap at the \evaluationEpisodes, we consider that the episode has been detected correctly. }
    \label{fig:evalMetric}\vspace{0.1in}
\end{figure}
\subsection{Evaluation Criteria and Metric.}
\emph{Eating Activity detection}: We evaluate the eating activity detection at two levels: (a) evaluate the possibility of detecting eating at a \textit{\evaluationSeconds}, and (b) evaluate the possibility of detecting the overall eating at a \textit{\evaluationEpisodes}.  
Figure~\ref{fig:evalMetric} pictorially describes the two levels.
For each level, we compute the \textit{precision}, \textit{recall} and \textit{F1-score}. 
A high precision value indicates that the seconds (or episodes) identified by \name as eating were actually seconds when eating occurred, whereas a high recall indicates that \name could find most of the moments (either \evaluationSecond or \evaluationEpisodes) when a participant performed an eating action. 
The \textit{F1-score} presents the harmonic mean between the precision and recall. 
We evaluated \name by performing a \ac{LOSOCV} and reporting the average performance of \names, which is the average of every participant's precision or recall or F1-score. 
The output of the \ac{GBM} classifier is utilized for the \evaluationSecond evaluation, while the output of the DBSCAN is utilized for the \evaluationEpisode evaluation. 
If there was 50\% overlap between the time of the predicted episode and the actual episode, we considered the episode as a True-Positive episode. 

\textit{Battery Lifetime}: Since the \nameG operates in several modes {(e.g., time-keeping mode, data collection mode, data logging mode)}, it is necessary to monitor the power consumption in each mode. Additionally, since our goal is to ensure that the \nameG can operate without frequent charging, it is necessary to understand the average battery lifetime. We measure the power consumption of each mode of operation in milliWatts and the average battery lifetime in terms of number of hours for which the device can operate after it is fully charged, until it is completely discharged. 

Now that we have established the evaluation criteria and metric, we next evaluate the performance of \names.

\subsection{$\mathsf{Q1:}$ Eating Activity Detection}
Although several previous studies have demonstrated the feasibility of automatically detecting eating in laboratory settings, very few researchers have explored the possibility of detecting eating over multiple days in free-living conditions, and evaluating their system at a \evaluationSeconds. We thus conducted the \studya to evaluate this possibility. 
We performed a leave-one-subject-out cross validation to generate the classification model as well as to determine the DBSCAN clustering parameters. 
At a \evaluationSecond analysis, the system attained a F1-score of 76.2\%. 
At \evaluationEpisode analysis, we observed that among the 76 meals that were consumed by the participants in the \studyas, we could correctly detect 63 meals. The system's average precision, recall, and F1-score across the participants during this \studya were 80.8\%, 86.8\%, and 81.6\% respectively. These results indicated that it was indeed possible to detect most meals that the individual consumes, even in naturalistic settings. 

\begin{figure}
    \centering
    \begin{subfigure}[b]{0.45\textwidth}
        \includegraphics[width=0.9\textwidth]{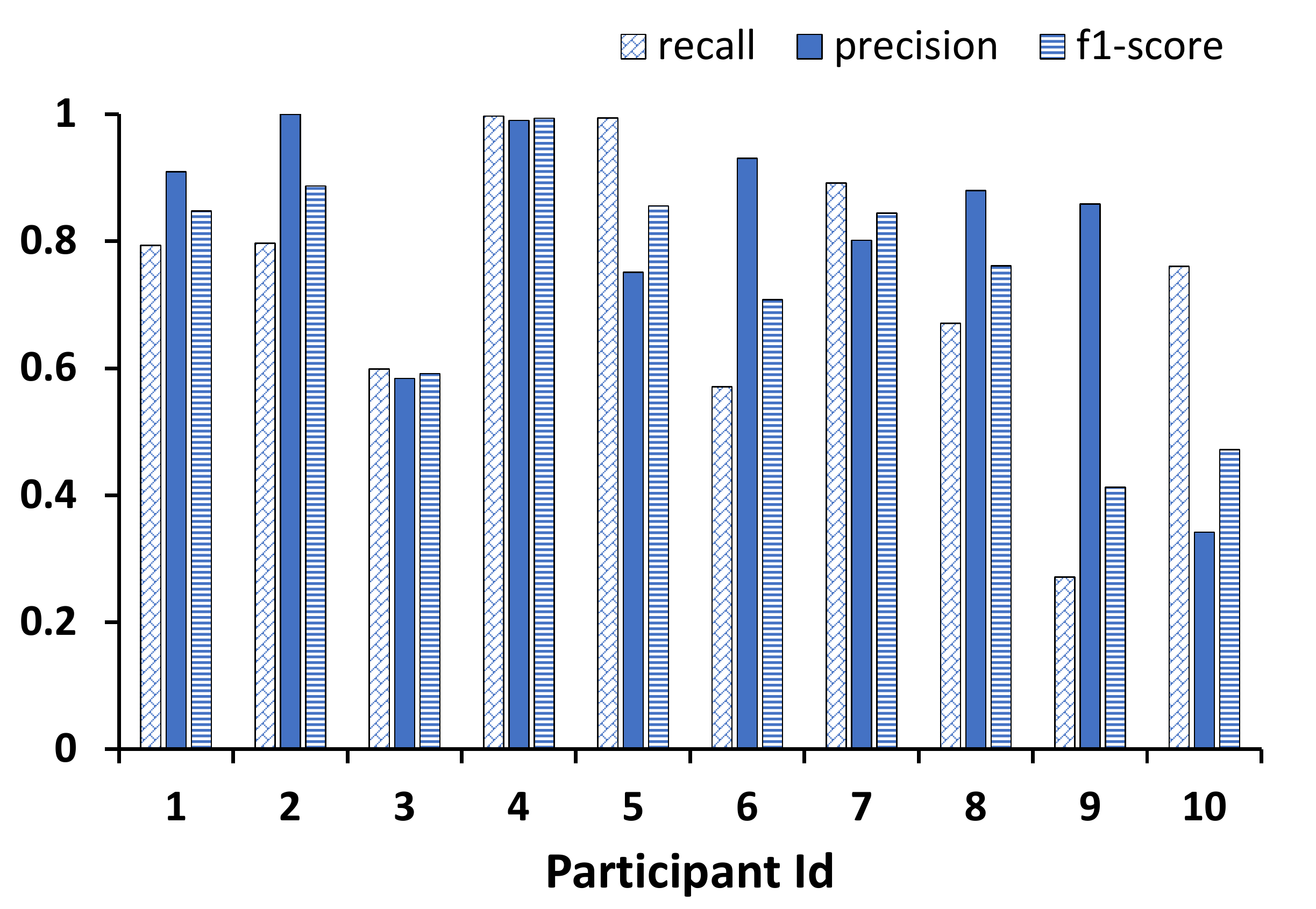}
        \caption{\evaluationEpisode}
        \label{fig:gull}
    \end{subfigure}
    \hfill
    \begin{subfigure}[b]{0.45\textwidth}
        \includegraphics[width=0.9\textwidth]{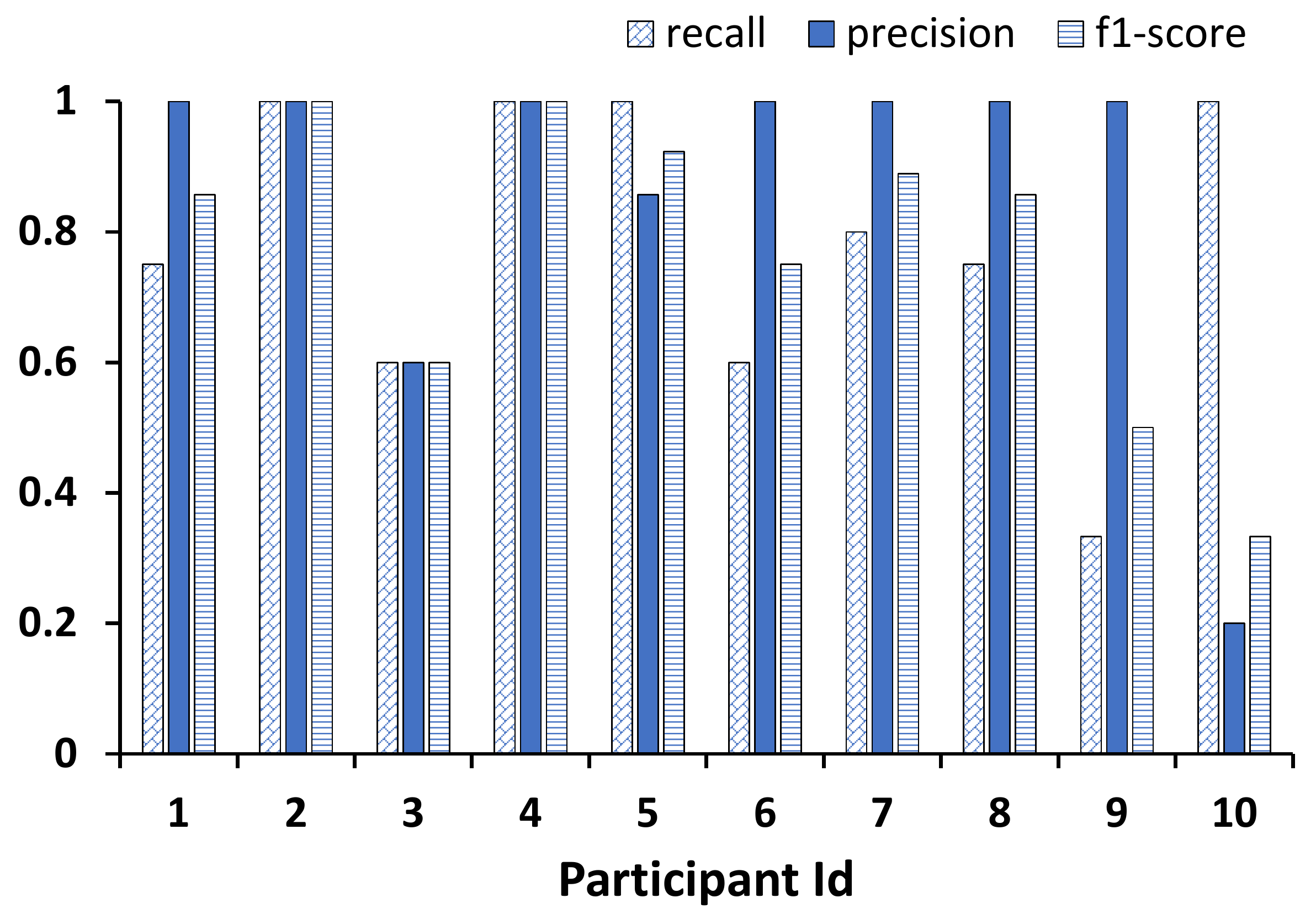}
        \caption{\evaluationSecond}
        \label{fig:tiger}
    \end{subfigure}
    \caption{Eating episode prediction evaluation for the \studyb using LOSOCV method for \textit{\evaluationSecond} and \textit{\evaluationEpisode} analysis. }\label{fig:chewing_second_result_afterremove}
\end{figure}

Through the \studya we demonstrated that the \nameG could indeed determine eating in a semi-free-living condition. To understand the system's performance in a completely uncontrolled setting, we analyzed the data from the \studybs. 
Overall, we found that at a \evaluationSeconds, our system could detect eating with an average F1-score of 73.7\% (average precision=80.5\%, average recall=73.4\%), while at a \evaluationEpisodes, the F1-score was 77.1\% (average precision=86.6\%, average recall=78.3\%). 
Figure~\ref{fig:chewing_second_result_afterremove} presents the per-participant performance of \names. 
From the result we can see that although our system performs well in semi-free-living settings, the performance degrades while evaluating in an uncontrolled free-living setting. 
This observation should motivate researchers to evaluate their systems not only in semi-controlled free-living condition, but in truly free-living conditions to identify their system's actual performance.

\noindpar{Comparison to other work.} Recently Chun et al. developed a neck-worn device for eating detection~\cite{Chun2018}. 
This device was deployed on only students for a single-day study (average 4.63 hours a person), without validation from a wearable camera. 
They also performed segmentation on a proximity signal using a threshold-based algorithm. 
Since we are interested in understanding how the two systems fare in determining chewing, the building block of an eating detection system, we compared their approach for segmentation recall against our approach in the \studya and observed a 13.1\% segmentation recall improvement in our approach as compared to their technique.

\subsection{$\mathsf{Q2:}$ Effect of Various Factors}
\noindpar{Performance of various sensors.}
To understand the usefulness of each sensor in determining the chewing sequence and the eating activity, we analyzed the \studyas's data and observed that for the \evaluationEpisode evaluation, we could achieve an average precision of 77.2\%, recall of 74.0\%, and F1-score of 73.4\% across participants when we use only the proximity sensor's signal, as compared to an average precision, recall and F1-score of 80.8\%, 86.8\%, and 81.6\% when we use all sensors. 
This improvement in performance validates the usefulness of employing multi-sensor modalities over using a single sensing modality, proximity, for eating detection. 


\noindpar{Feature Importance.}
\label{sec:featureselection}
Since the number of instances for each feature in the gradient boosting tree is proportional to the feature's contribution among all the features, we ranked the features accordingly. 
In Section~\ref{sec:featExtract} we described the features were extracted from either the Bite Window ($BW$) or the Chewing Window ($CW$). 
We observed that the top-5 features that aided in classification, as selected by the gradient boosting algorithm were: (1) Frequency: FFT 2.5 Hz of energy signal ($BW$), (2) Time-series: first location of minimum of energy signal ($CW$), (3) Time-series: first location of maximum of energy signal ($CW$), (4) Frequency:FFT 0.5 Hz of ambient signal ($CW$), and (5) Time-series: Count above mean of ambient signal ($CW$). 
All top-5 features were extracted from both the ambient light and energy signals. Two among the five features were FFT-based features, while three were time-series-based features. Four features were extracted from the $CW$, and one was from the $BW$.

\begin{figure}[t]
\centering
\includegraphics[width=0.9\columnwidth]{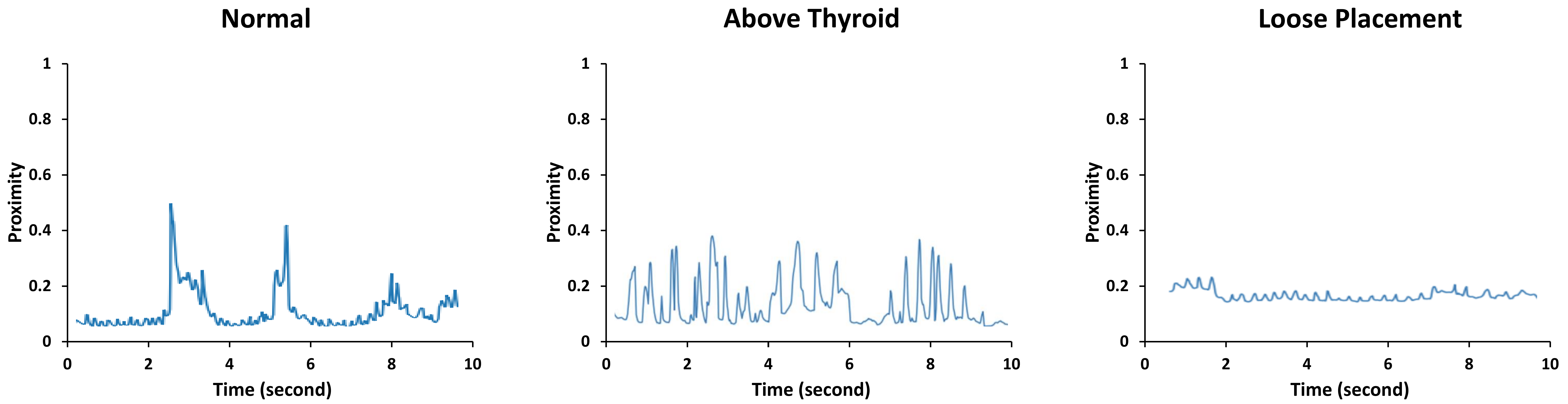}\vspace{-0.15in}
\caption{Proximity sensor signals of different sensor positions.
}
\label{fig:edge_case}
\end{figure}

\noindpar{Effect of various \nameG positions.}
The necklace form factor, worn loosely around the neck, lets participants adjust and move the necklace closer or farther from their mouth.
Using the ground truth data, we documented various placements of the necklace for our study participants (normal, above the thyroid, and loosely bound). Figure~\ref{fig:edge_case} presents the signal collected for these cases while a participant consumed rice crackers. In all the cases, even though the proximity signal spanned different amplitude ranges, periodic prominent peaks were still observed in the signal, allowing us to predict chewing sequences.

\subsection{$\mathsf{Q3:}$ Battery Lifetime}
\begin{figure}[t]
\centering
\includegraphics[width=80mm]{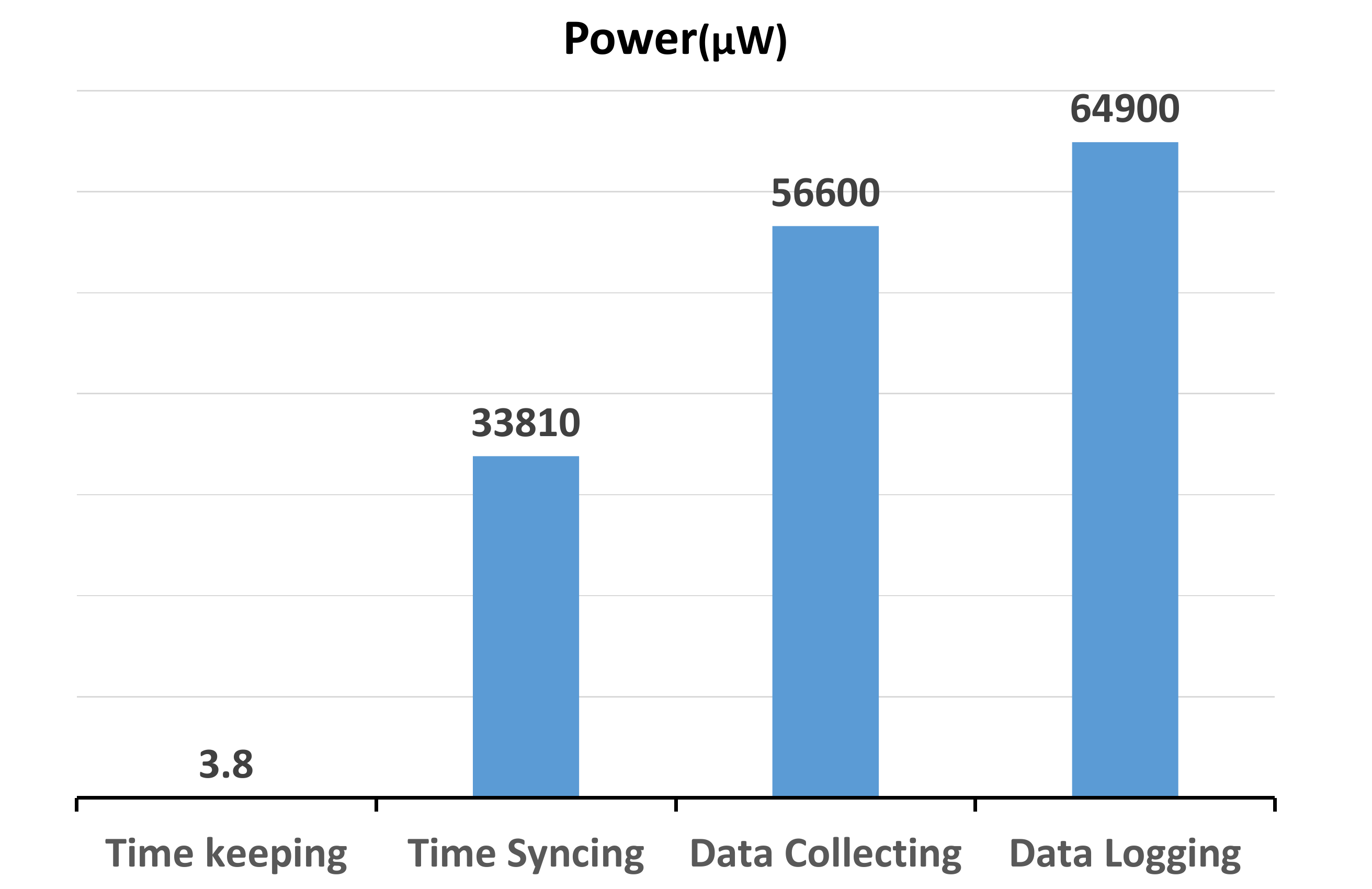}
\caption{Power consumption for each active mode.}
\label{fig:battery_life}
\end{figure}

Now that we have established that \name can indeed detect eating activities in naturalistic setting, we next analyse whether the device can collect sensor data continuously during an individual's waking hours. 
The \nameG is powered by a 350mAh battery to ensure a relatively small size. 
To conserve power we implemented several power saving schemes. Firstly, we selected the Nordic NRF52832 SiP that includes a low powered microcontroller and a BLE interface. Overall, this choice helped in reducing power consumption as well as the size. Secondly, we identified that the SD card writes were power hungry. We thus implemented a batch and write approach for data logging. Thirdly, we designed the super low power time keeping mode with extra tiny battery so we can always have a valid system time even the main battery is off. 
The \nameG can operate in four modes -- time keeping mode ({only running the RTC}), time syncing mode ({communicating over BLE to synchronize time }), data collecting mode ({recording sensor reading}) and data logging mode ({writing data to SD card}). 
We measured the average power for each mode using the STM32 Nucleo expansion board~\cite{STM32} for power measurement, and tested the battery life time for a single charge. 
The power consumption for each mode, as shown in Figure~\ref{fig:battery_life}, were 3.8$\mu$W, 36.8 mW, 56.6 mW, and 64.9 mW  for time keeping, time syncing, data collecting and data logging mode respectively. 
Overall, in the \studya the average battery life was 13 hours and  it improved to 15.8 hours in the \studybs, which is sufficient to record data for most of the day time and cover all the meals that occur during a waking-day. 
This lifetime allows for simplified deployments and little, if any, study coordinator oversight. 
The participants can charge the device once a day before sleep. 

\subsection{Summary of Results}
Based on these results, we observe that: (i) it is indeed possible to detect the eating activity in a \textit{completely naturalistic setting} at an \textit{extremely challenging per-second granularity} with a F1-score of 73.7\%  (76.2\% in semi-free living conditions), thus presenting the possibility of performing real-time interventions. This is an improvement over similar existing technique. (ii) It is possible to determine the eating episodes, even as an individual perform their everyday activities, with a F1-score of 77.1\% (81.6\% in semi-free living conditions). (iii) A combination of multi-sensor eating detection outperforms the eating detection by a single sensing modality. (iv) Once \namef battery is completely charged, it can continuously monitor a participant's eating activity in free living conditions for over 15 hours, thus making it possible to monitoring individuals through an entire waking-day. 
These results demonstrate that the \nameG can be a promising avenue for identifying eating-related activities through an individual's entire waking day, and deploying and testing eating-related interventions in real-time.

Overall, our results are promising. However, given that both the studies was conducted in a truly free-living population setting, challenging eating postures and scenarios confounded some of our chewing and eating episode detection. For example, lying down while eating confounded our ability to capture the lean forward motion during eating and also the chewing from the proximity sensor. 
Similarly, eating in total darkness confounded the classification model because the ambient light sensor gave readings far below the normal range, while eating during exercising resulted in several false chewing detection, although our aggregate filter successfully filtered out these isolated chewing sequences. Meals that included no chewing also presented a challenge; this was seen in one of our participants who had 50\% of the labeled eating episodes as ice-cream, yogurt, or other non-chewing foods. This issue has also been reported by other researchers exploring chewing based activity detection~\cite{bi:ubicomp18}. In future we will explore techniques to overcome these challenges. 

%% file: 2_related.tex
\section{Related Work}
\label{sec:related}
Large-scale, whole population interventions~\cite{marcus1998physical}, such as ad-campaigns targeted towards curtailing eating-related disorders (e.g., obesity), have had little or no success in addressing the epidemic~\cite{pember2017dietary,neuhauser2010ehealth,neuhauser2003rethinking,tufano2005mobile}.
Instead, researchers are proposing just-in-time interventions~\cite{nahum2014just,farooq2017reduction,nahum2015building,jaimes2015trends} to test personalized interventions that are tailored to person-specific needs
~\cite{ryan2002efficacy,tang2015can}. 
Detecting eating automatically is the first step towards testing these personalized just-in-time intervention. 
However, several many factors make automated eating detection challenging to implement. These factors range from identifying the right device, signals, or form-factor, to validating eating activity using visual confirmation in real-world settings.

\subsection{Eating Detection Techniques} 
Researchers have proposed several techniques to automatically detect the eating activity. 
We detail some of the proposed techniques, grouped based on signal type. 

\noindpar{Audio and Video}: Automated sensing platforms using image- and audio-based techniques with sensors placed around the throat or ear have shown promise in detecting eating~\cite{amft2006emgsound,rahman2014bodybeat,nishimura2008eating,bi:ubicomp18, ThomazImage2013, ThomazAmbient2015, mirtchouk2017ubicomp}. However, the utility of these sensors is limited by short battery life (reducing autonomy), and security or privacy concerns. Eating detection systems designed without camera or audio components (as in this work) reduce privacy concerns and enable longer battery life.

\noindpar{Physiological:} Various techniques have been used to indirectly detect chewing, which involve \ac{EMG}~\cite{zhang2017eyeglass} sensors on an eye-glass frame, in-the-ear microphones combined with photoplethysmography (PPG) sensors~\cite{papapanagiotou2017ppg}, and mechanoelectrical sensors placed directly around the neck~\cite{amft2006emgsound}, among others~\cite{kalantarian2014wearable,kalantarian2015necklace}.
While these sensors have shown promise in controlled environments, they have not been tested in free-living populations, over significant periods of time, and many need direct skin contact, which can be uncomfortable and affect adherence, thereby limiting their potential utility in longitudinal studies.

\noindpar{Inertial or Proximity:} Several researchers have proposed techniques to automatically detect eating using inertial sensors embedded in wrist-worn devices and phones~\cite{Thomaz2015, Dong2014, mirtchouk2017ubicomp, sen2015case}. 
However, wrist-worn sensors are limited by several confounding gestures \cite{HASCA2017whenFail}. 
More recently, researchers have explored the possibility of detecting chewing bouts and eating episodes using only an on-neck proximity sensor, combined with a threshold-based algorithm~\cite{Chun2018}. We show that our multi-sensor method outperforms such a threshold-based method in recalling chewing sequences, even in naturalistic setting.  

Our method re-imagines \textbf{inertial or proximity} sensing modalities by fusing them with data from an ambient light sensor, chewing-related periodic subsequence features, and time of day. 
We present an eating detection framework that uses this fusion and is tested in free-living settings. 

\begin{table*}[t]
\centering 
\begin{small}
\begin{tabular*}{\textwidth}{p{0.5cm}p{2.55cm}p{1.8cm}p{2.1cm}cc ccc}

\toprule
    {Year} & {Study} & {Sensors} & {\stackon{position}{On-body}}
    & \mc{\stackon{participants}{No. of}} & \mc{\stackon{per day}{Avg hours}} &\mc{\stackon{video}{Validation}}
    & \mc{\stackon{student}{Non-}} & \mc{Obese}\\
    \midrule
2014  & Fontana et al.~\cite{fontana2014automatic} & S1, S4, S6	 & Ear, Wrist, Chest & 12 & 24.0 &  \xmark & \cmark & \cmark\\
2015& {Thomaz et al.~\cite{Thomaz2015}}&S1&Wrist&7+1&5.7 / 13.6&\xmark&\xmark&\xmark\\
2015 & Bedri et al.~\cite{bedri2015detectingmastication} & S2, S5	& Ear, Head & 6 & 6.0  & \xmark & \cmark & \xmark\\
2016  & Farooq et al.~\cite{farooq2016chewing}  & S4   & Temple  & 8  & 3.0	  & \xmark  & \cmark  & \xmark\\
2017 & Bedri et al.~\cite{bedri2017earbit} & S1-S3, S5, S7 &	Neck, Ear & 	10 & 4.5 & \cmark & \cmark & \xmark\\
2017 & Zhang et al.~\cite{zhang2017eyeglass} & S8 & Ear & 10 & 6.1 & \xmark	 & \xmark	 & \xmark\\
2017 & 	Mirtchouk et al.~\cite{mirtchouk2017ubicomp} & S1-S3, S7 &	Ear, Wrist, Head  & 11 & 	11.7 & 	\xmark	 & \cmark	 & \xmark\\
2018 & Sen et al.~\cite{sen2018annapurna}&S1, S2, S10& Wrist& 9& 5.8&\xmark&\cmark&\xmark\\
2018 & Chun et al.~\cite{Chun2018} & S5 & Neck & 17 & 4.6 & \xmark & \xmark  & 	\xmark \\
2018 & Bi et al.~\cite{bi:ubicomp18}	 & S7	&Ear	 & 14  & 2.3  & \cmark	 & \xmark & \xmark	 \\
2019 & This Work& S1-S3, S5, S9  & Neck & 10+10 & 4.9/9.5 & \cmark & \cmark & \cmark\\
\bottomrule

\end{tabular*}
\end{small}
\begin{captiontext}
\begin{tiny}
\RaggedRight S1 - accelerometer, S2 - gyroscope, S3 - magnetometer, S4 - piezo, S5 - proximity, S6 - RF, S7 - microphone, S8 - EMG, S9 - light, S10 - camera
\end{tiny}
\end{captiontext}
\caption{Comparing the literature on in-wild eating detection to this work. }
\label{tbl:eating_detect_table}
\end{table*}

\subsection{Eating Detection in Naturalistic Setting}
Researcher have explored the use of various sensors for detecting the eating activity outside laboratory settings~\cite{zhang2017eyeglass,papapanagiotou2017ppg,bedri2015detectingmastication,bedri2017earbit,farooq2016chewing, farooq2016detection,fontana2014automatic}, but the length of continuously recorded experimentation rarely exceeds more than few hours in a day, limiting its potential for longitudinal eating-related studies. 
Table~\ref{tbl:eating_detect_table} lists some existing research that utilizes various sensing modalities to detect the eating activity.  
Overall, we identified some key factors have made these free-living studies hard to execute:

\noindpar{Validation Requirement:} A straightforward validation technique that has been adopted by researchers is self-reporting, i.e., the researchers requested participants to note down the start and end time of every meal that the participant consumes~\cite{Dong2012,Scisco2014,sen2018annapurna} or create a food journal~\cite{papapanagiotou2017ppg, zhang2017eyeglass}. 
However, manually noting details about every meal can be burdensome and error-prone~\cite{Heitmann1995}. 
To reduce the burden on participants, several researchers have proposed the use of a front-, upward- or downward- facing camera~\cite{Yatani2012, ThomazImage2013, hodges2006sensecam, Thomaz2015, bedri2017earbit,bi:ubicomp18} for validating their sensor's or device's performance.
Researchers have tested a camera around the chest, shoulder, and wrist and found the shoulder camera to be best for privacy, and easiest to wear~\cite{CHI2017WillSenseRawan}. 
This placement was also found to be best at capturing eating related activities~\cite{PercomInvestigatingBarrier}. 
We thus use a shoulder mounted camera for validating our device's performance.

\noindpar{Diverse Population Requirement:} While several sensing systems have shown promise in the wild, they have predominantly been validated within the student population\hey{~\cite{bi:ubicomp18,Chun2018}}. Before such systems can be truly generalizable, additional data is needed from a diverse population sample (especially including people belonging to various body mass ranges) that is not only student-based or healthy people focused. People with varied body shapes may experience varied comfort levels and accuracy of any eating detection system, potentially confounding the system, but enabling deeper insight and translation of research to practice. 
To the best of our knowledge, we are among the few to explicitly validate our automated eating detection systems with an obese population group.



%% file: 8_conclusions.tex
\section{Conclusion}
In this paper we present the design, implementation, and evaluation of a \nameG suite for detection and validation of chewing sequences and eating episodes in free living conditions. 
We utilize sensor data from a proximity sensor, an ambient light sensor and an \ac{IMU} sensor to detect chewing activity. 
We performed two free-living studies using the \nameGs. In our exploratory study, where selective meals are captured and non-continuous wear time is observed, we are able to outperform other methods with a 76.2\% F1-score at the \evaluationSeconds, and 81.6\% at the \evaluationEpisodes.
Overall, our system achieves a F1-score of 73.7\% in detecting the chewing sequences at an extremely challenging per-second level granularity in a truly free-living study. Additionally, our system could detect eating episodes with a F1-score of 77.1\%. Additionally, the \nameG could monitor the participants for 15.8 hours during the waking-day, making it possible to monitor an individual's eating activity occurring during an entire day. 

Our dataset is unique in that it comprises a 2-week and 2-day study of eating detection collected from 20 participants in free living condition. This data is accompanied by video that is validated and professionally labeled, marking the ground truth. 
This dataset includes many unique eating events and eating situations including eating in cars, slouching, talking and eating, and eating a variety of soft foods. 
We hope this dataset will provide a repository from which researchers can glean insights from the data and inform future studies. 
Understanding the effects of chewing speed and duration on overeating can help interventionists design and test improved treatments for eating behavior monitoring.
\name is the next step in enabling behaviorists to imagine, design, deploy, and test for effective interventions during and immediately after an eating episode.

